\documentclass[longauth]{aa}
\usepackage{graphicx}
\usepackage{float}
\usepackage{amssymb}
\usepackage{amsmath}
\usepackage{color}
\usepackage{soul}
\usepackage{multirow}
\usepackage[T1]{fontenc}
\usepackage[draft]{hyperref}
\graphicspath{{./figures/}}

\usepackage{babel}

\newcommand{\sarc}{$^{\prime\prime}\!\!\,$}

\newcommand{\ddfp}{\textsc{ddf-pipeline}}
\newcommand{\ndppp}{\textsc{DPPP}}
\newcommand{\prefac}{\textsc{prefactor}}
\newcommand{\difmap}{\textsc{difmap}}
\newcommand{\wsclean}{\textsc{wsclean}}
\newcommand{\lofvlbi}{\textsc{LOFAR-VLBI}}

\definecolor{Mygrey}{gray}{0.75}

\begin{document}
\title{Sub-arcsecond imaging with the International LOFAR Telescope}
\subtitle{I. Foundational calibration strategy and pipeline}
\titlerunning{Sub-arcsecond LOFAR}
\author{L.~K. Morabito\inst{1,2}\thanks{E-mail: leah.k.morabito@durham.ac.uk}
\and N.~J. Jackson\inst{3}
\and S. Mooney\inst{4}
\and F. Sweijen\inst{5}
\and S. Badole\inst{3}
\and P. Kukreti\inst{6,7}
\and D. Venkattu\inst{8}
\and C. Groeneveld\inst{5}
\and A. Kappes\inst{9}
\and E. Bonnassieux\inst{10}
\and A. Drabent\inst{11}
\and M. Iacobelli\inst{7}
\and J.~H. Croston\inst{12}
\and P.~N. Best\inst{13}
\and M. Bondi\inst{14}
\and J.~R. Callingham\inst{5,7}
\and J.~E. Conway\inst{15}
\and A.~T. Deller\inst{16}
\and M.~J. Hardcastle\inst{17}
\and J.~P. McKean\inst{7,6}
\and G.~K. Miley\inst{5}
\and J. Moldon\inst{18}
\and H.~J.~A. R\"ottgering\inst{5}
\and C. Tasse\inst{19,20}
\and T.~W. Shimwell\inst{7,5}
\and R.~J.~van Weeren\inst{5}
\and J.~M. Anderson\inst{21,22}
\and A. Asgekar\inst{7,23}
\and I.~M. Avruch\inst{7,24}
\and I.~M.~van Bemmel\inst{25}
\and M.~J. Bentum\inst{7,26}
\and A. Bonafede\inst{27,14,28}
\and W.~N. Brouw\inst{6}
\and H.~R. Butcher\inst{29}
\and B. Ciardi\inst{30}
\and A. Corstanje\inst{31,32}
\and A. Coolen\inst{7}
\and S. Damstra\inst{7}
\and F.~de Gasperin\inst{28,14}
\and S. Duscha\inst{7}
\and J. Eisl\"offel\inst{11}
\and D. Engels\inst{28}
\and H. Falcke\inst{32}
\and M.~A. Garrett\inst{3,5}
\and J. Griessmeier\inst{33,34}
\and A.~W. Gunst\inst{7}
\and M.~P.~van Haarlem\inst{7}
\and M. Hoeft\inst{11}
\and A.~J.~van~der Horst\inst{35,36}
\and E. J\"utte\inst{37}
\and M. Kadler\inst{9}
\and L.~V.~E. Koopmans\inst{6}
\and A. Krankowski\inst{38}
\and G. Mann\inst{39}
\and A. Nelles\inst{40}
\and J.~B.~R. Oonk\inst{41}
\and E. Orru\inst{7}
\and H. Paas\inst{42}
\and V.~N. Pandey\inst{7}
\and R.~F. Pizzo\inst{7}
\and M. Pandey-Pommier\inst{43}
\and W. Reich\inst{44}
\and H. Rothkaehl\inst{45}
\and M. Ruiter\inst{7}
\and D.~J. Schwarz\inst{46}
\and A. Shulevski\inst{5,47}
\and M. Soida\inst{48}
\and M. Tagger\inst{33}
\and C. Vocks\inst{39}
\and R.~A.~M.~J. Wijers\inst{47}
\and S.~J. Wijnholds\inst{7}
\and O. Wucknitz\inst{44}
\and P. Zarka\inst{49,34}
\and P. Zucca\inst{7}
}
\authorrunning{L.K. Morabito et al.}
\institute{ Centre for Extragalactic Astronomy, Department of Physics, Durham University, Durham DH1 3LE, UK
\and Institute for Computational Cosmology, Department of Physics, University of Durham, South Road, Durham DH1 3LE, UK
\and Jodrell Bank Centre for Astrophysics, Department of Physics and Astronomy, School of Natural Sciences, University of Manchester, Manchester M13 9PL
\and School of Physics, University College Dublin, Belfield, Dublin 4, Ireland
\and Leiden Observatory, Leiden University, P.O. Box 9513, 2300 RA Leiden, The Netherlands
\and Kapteyn Astronomical Institute, University of Groningen, Postbus 800, 9700 AV Groningen, The Netherlands
\and ASTRON, Netherlands Institute for Radio Astronomy, Oude Hoogeveensedijk 4, 7991 PD, Dwingeloo, the Netherlands
\and Department of Astronomy, AlbaNova University Center, Stockholm University, SE-10691 Stockholm, Sweden
\and Institut f\"ur Theoretische Physik und Astrophysik, Universit\"at W\"urzburg, Emil-Fischer-Str. 31, 97074 W\"urzburg, Germany
\and Universita di Bologna, Via Zamboni, 33, 40126 Bologna BO, Italy
\and Th\"uringer Landessternwarte, Sternwarte 5, D-07778 Tautenburg, Germany
\and School of Physical Sciences, The Open University, Walton Hall, Milton Keynes, MK7 6AA, UK
\and Institute for Astronomy, University of Edinburgh, Royal Observatory, Blackford Hill, Edinburgh, EH9 3HJ, UK
\and INAF - Istituto di Radioastronomia, Via P. Gobetti 101, 40129, Bologna, Italy
\and Department of Space, Earth and Environment, Chalmers University of Technology, Onsala Space Observatory, SE-439 92 Onsala, Sweden
\and Centre for Astrophysics and Supercomputing, Swinburne University of Technology, Mail H30, PO Box 218, Hawthorn, VIC 3122,  Australia
\and Centre for Astrophysics Research, University of Hertfordshire, College Lane, Hatfield AL10 9AB, UK
\and Instituto de Astrof\'isica de Andaluc\'ia (IAA, CSIC), Glorieta de las Astronom\'ia, s/n, E-18008 Granada, Spain
\and GEPI \& USN, Observatoire de Paris, CNRS, Universit\'e Paris Diderot, 5 place Jules Janssen, 92190 Meudon, France Grahamstown 6140, South Africa
\and Centre for Radio Astronomy Techniques and Technologies, Department of Physics and Electronics, Rhodes University, Grahamstown 6140, South Africa
\and Technische Universit\"at Berlin, Institut f\"ur Geod\"asie und Geoinformationstechnik, Fakult\"t VI, Sekr. KAI 2-2, Kaiserin-Augusta-Allee 104-106, D-10553 Berlin, Germany
\and GFZ German Research Centre for Geosciences, Telegrafenberg, D-14473 Potsdam, Germany
\and Currently at Shell Technology Center, Bangalore, India 562149
\and Science and Technology B.V., Delft, the Netherlands
\and Joint Institute for VLBI ERIC (JIVE), Oude Hoogeveensedijk 4, 7991 PD Dwingeloo, The Netherlands
\and Eindhoven University of Technology, De Rondom 70, 5612 AP Eindhoven, The Netherlands
\and DIFA -- Universit\'a di Bologna, via Gobetti 93/2, I-40129 Bologna, Italy
\and Hamburger Sternwarte, Universit\"at Hamburg, Gojenbergsweg 112, D-21029, Hamburg, Germany
\and Research School of Astronomy \& Astrophysics, Mt. Stromlo Observatory, Cotter Road, Weston Creek, ACT 2611 AUSTRALIA
\and Max-Planck Institute for Astrophysics, Karl-Schwarzschild-Strasse 1. 85748, Garching, Germany
\and Astrophysical Institute, Vrije Universiteit Brussel, Pleinlaan 2, 1050 Brussels, Belgium
\and Department of Astrophysics/IMAPP, Radboud University Nijmegen, P.O. Box 9010, 6500 GL Nijmegen, The Netherlands
\and LPC2E -- Universit\'e d'Orl\'eans  / CNRS, 45071 Orl\'{e}ans cedex 2, France
\and Station de Radioastronomie de Nan\c{c}ay, Observatoire de Paris, PSL Research University, CNRS, Univ. Orl\'{e}ans, OSUC, 18330 Nan\c{c}ay, France
\and Department of Physics, The George Washington University, 725 21st Street NW, Washington, DC 20052, USA
\and Astronomy, Physics and Statistics Institute of Sciences (APSIS), The George Washington University, Washington, DC 20052, USA
\and Ruhr University Bochum, Faculty of Physics and Astronomy, Astronomical Institute, 44780 Bochum, Germany
\and Space Radio-Diagnostics Research Centre, University of Warmia and Mazury, ul. Romana Prawochenskiego 9, 10-719 Olsztyn, Poland
\and Leibniz-Institut f\"ur Astrophysik Potsdam (AIP), An der Sternwarte 16, 14482 Potsdam, Germany
\and ECAP, Friedrich-Alexander-University Erlangen-Nuremberg, Erwin-Rommel-Str. 1, 91058 Erlangen, Germany   DESY, Platanenallee 6, 15738 Zeuthen, Germany
\and SURF/SURFsara, P.O. Box 94613, 1090 GP Amsterdam, The Netherlands
\and CIT, Rijksuniversiteit Groningen, Nettelbosje 1, 9747AJ Groningen, The Netherlands
\and Universit\'e Claude Bernard Lyon1, Ens de Lyon, CNRS, Centre de Recherche Astrophysique de Lyon, 43 Boulevard du 11 Novembre 1918, 69100 Villeurbanne, France
\and Max-Planck-Institut f\"ur Radioastronomie, Auf dem H\"ugel 69, 53121 Bonn, Germany
\and CBK PAN , Bartycka 18 A,00-716 Warsaw, Poland
\and Fakult\"at f\"ur Physik, Universit\"at Bielefeld, Postfach 100131, 33501, Bielefeld, Germany
\and Anton Pannekoek Institute for Astronomy, University of Amsterdam, Postbus 94249, 1090 GE Amsterdam, The Netherlands
\and Astronomical Observatory of the Jagiellonian University, ul. Orla171, 30-244 Krak\'ow, Poland
\and LESIA, Observatoire de Paris, CNRS, PSL, SU, UP, Place J. Janssen, 92190 Meudon, France }

\date{Received XXX; accepted YYY}

\abstract{The International LOFAR Telescope is an interferometer with stations spread across Europe. With baselines of up to $\sim$ 2,000 km, LOFAR has the unique capability of achieving sub-arcsecond resolution at frequencies below 200$\,$MHz. However, it is technically and logistically challenging to process LOFAR data at this resolution. To date only a handful of publications have exploited this capability. 
Here we present a calibration strategy that builds on previous high-resolution work with LOFAR. It is implemented in a pipeline using mostly dedicated LOFAR software tools and the same processing framework as the LOFAR Two-metre Sky Survey (LoTSS). We give an overview of the calibration strategy and discuss the special challenges inherent to enacting high-resolution imaging with LOFAR, and describe the pipeline, which is publicly available, in detail. 
We demonstrate the calibration strategy by using the pipeline on P205+55, a typical LoTSS pointing with an 8 hour observation and 13 international stations. We perform in-field delay calibration, solution referencing to other calibrators in the field, self-calibration of these calibrators, and imaging of example directions of interest in the field. 
We find that for this specific field and these ionospheric conditions, dispersive delay solutions can be transferred between calibrators up to $\sim$1.5 degrees away, while phase solution transferral works well over $\sim$1 degree. 
We also demonstrate a check of the astrometry and flux density scale with the in-field delay calibrator source. 
Imaging in 17 directions, we find the restoring beam is typically $\sim$0.3\sarc $\times$0.2\sarc\ although this varies slightly over the entire 5 square degree field of view. We find we can achieve $\sim$80 to 300$\,\mu$Jy$\,$bm$^{-1}$ image rms noise, which is dependent on the distance from the phase centre; typical values are $\sim 90\,\mu$Jy$\,$bm$^{-1}$ for the 8 hour observation with 48 MHz of bandwidth.  Seventy percent of processed sources are detected, and from this we estimate that we should be able to image roughly 900 sources per LoTSS pointing. This equates to $\sim$3 million sources in the northern sky, which LoTSS will entirely cover in the next several years. Future optimisation of the calibration strategy for efficient post-processing of LoTSS at high resolution (LoTSS-HR) makes this estimate a lower limit. }

\maketitle

\section{Introduction}
\label{sec:introduction}
The LOw-Frequency Array \citep[LOFAR;][]{van_haarlem_lofar_2013} is an interferometer that operates at frequencies between 10 and 240 MHz. The facility currently consists of 52 stations spread throughout Europe. Thirty-eight of these stations are located in the Netherlands: 24 `core' stations within 4$\,$km of the array centre in Exloo and a further 14 `remote' stations with baselines of up to 120$\,$km. Calibrating the stations in the Netherlands to achieve 6\sarc\ \ resolution maps in a 20 square degree field of view at 150$\,$MHz is now relatively routine due to the development of calibration and imaging strategies that cope with direction-dependent effects (DDEs) at low frequencies \citep{van_weeren_lofar_2016,tasse_faceting_2017,tasse_lofar_2020}. There are almost 400 publications to date that feature LOFAR results, which span a wide range of topics. However, the vast majority of these use only stations in the Netherlands, for imaging, polarisation, or time domain observations. 

The LOFAR array has been expanded now to include 14 `international' stations, six of which are in Germany (Effelsberg, Unterweilenbach, Tautenburg, J\"{u}lich, Potsdam, and Norderstedt), three are in Poland (Bor\'{o}wiec, \L azy, and Baldy), and one each is in the UK (Chilbolton), France (Nan\c{c}ay), Sweden (Onsala), Ireland (Birr), and Latvia (Ventspils). This geographic spread provides baselines of up to 1989$\,$km (from Birr to \L azy), which yields an angular resolution of 0.27\sarc\ \ at 150 MHz. New stations are also being planned. The first will become operative in Italy, which will improve the sensitivity, $u$-$v$ coverage, and resolution. No other low-frequency instrument is capable of such resolution, nor is such an instrument planned to exist, even in the era of the Square Kilometre Array (SKA). 

Opening up the megahertz frequency regime to high-resolution imaging is desirable for a wide range of science cases. This has been demonstrated with both the LOFAR high band antenna (HBA; 120--240$\,$MHz) and low band antenna (LBA; 30--78$\,$MHz) for individual sources. All but one of the published studies conducted with the international stations use the HBA \citep{moldon_lofar_2015,jackson_lbcs_2016,varenius_subarcsecond_2015,varenius_subarcsecond_2016,ramirez_subarcsecond_2018,harris_lofar_2019,kappes_lofar_2019}. Excellent examples of the scientific potential include a study of M82 by \cite{varenius_subarcsecond_2015}, where the authors investigated the properties of the radio spectra at 154$\,$MHz in the nuclear region of the nearby galaxy, detecting seven previously uncatalogued compact objects, including supernova remnants, and providing spatially resolved information for fitting the radio spectral energy distributions (SEDs) of the supernova remnants. There is also significant discovery potential with the LBA. For example \cite{morabito_lofar_2016} used observations centred on 54$\,$MHz to study the resolved morphological and spectral properties of a high-redshift radio galaxy, showing that they are similar to low-redshift sources of the same class; without the low-frequency high-resolution imaging, this would not have been possible. These scientific results all rely on using the full international LOFAR to resolve substructure at low frequencies where synchrotron processes dominate the radio spectrum. 

The small number of publications making use of the full international LOFAR telescope is due to the fact that the calibration is technically challenging and the data volumes are large. At low frequencies, the ionosphere plays a large part in corrupting astronomical signals \citep[for a detailed review see][]{intema_ionospheric_2009}. Even considering just the area over the stations in the Netherlands, which are all within 120$\,$km of one another, the ionospheric conditions can be different; the wide geographic spread of international LOFAR stations exacerbates this problem. Another issue is that although LOFAR is technically a connected interferometer (all stations have fibre feeds to the correlator), the international and remote stations have independent clocks. Offsets in the clock values (and, to a lesser extent, clock drifts) introduce delays in the phase structure of the data. Additionally, the correlator uses a geometric model to synchronise data from different stations, and the tolerance for deviations from this model decreases as the distance between stations increases. Finally, the international stations have different station beams than the core and remote stations. By default, the international stations use all 96 antenna tiles, meaning they have a larger geographic spread than the compact core and remote stations, which can only use 48 tiles simultaneously. 
However, the international stations have two main advantages. First, the international stations are twice as sensitive (providing $\sqrt{2}$ increased sensitivity on baselines containing them), which is important in the low signal-to-noise regime. Second, the larger effective area of the stations translates to smaller fields of view, which are also reduced by bandwidth and time smearing. Smaller fields of view mean extreme wide-field effects can largely be ignored. 

Another issue is calibration sources. Observing calibrator sources allows us to solve for effects that corrupt the data, including instrumental effects and sky (propagation) effects \citep[an excellent overview is given in ][]{smirnov_revisiting_2011}. Standard flux density calibrators are bright sources with high signal-to-noise and for which the flux density and source structure is well known at the resolution of the observations; we use this information to set the absolute flux density scale and remove direction-independent instrumental effects.  After this correction, we can use fainter secondary (or phase) calibrators to remove the effects of signal corruption in the direction of our target of interest. These effects will mainly corrupt the phases, but it is not uncommon to perform (slow) amplitude calibration on the secondary calibrators.

Even with appropriate calibrator sources, it can be difficult to calibrate the international LOFAR array to produce reliably high-fidelity, high-resolution images. The ionosphere introduces dispersive (i.e. frequency-dependent) delays: $d = \Delta \phi / \Delta \nu$, where $d$ is delay, $\phi$ is phase, and $\nu$ is frequency. At the LOFAR observing frequencies, this is a dominant effect. This behaviour means we cannot rely on delay calibration routines that assume that $d$ is independent of $\nu$, which is the case for virtually all radio calibration software\footnote{A basic fringe-fitting algorithm that solves for dispersive delays has only recently been added to the Common Astronomy Software Applications \citep[CASA;][]{mcmullin_casa_2007}, but this is still in the experimental stage.}. Source structure is another major challenge for proper calibration. With increasing resolution, the number of sources that have truly compact source structure drops drastically as extended emission drops out of the spatial filter of the longer baselines. This has several effects: \textit{(i)} the sky density of sources decreases; \textit{(ii)} there is less total flux on compact scales and we move into a lower signal-to-noise regime; and \textit{(iii)} resolved sources have increasingly complicated visibility information and suitable models are necessary to drive self-calibration to convergence. 

Despite these challenges, we have made a substantial amount of progress over the last few years. The Long Baseline Calibrator Survey \citep[LBCS;][]{jackson_lbcs_2021,jackson_lbcs_2016} is now complete and provides a database of  calibrator sources over the northern sky. We have developed a calibration strategy and built a pipeline to carry out this strategy, which forms the basis of high-resolution imaging with LOFAR. Additionally, initial tests have shown that we are able to image targets within LOFAR's wide field of view at sub-arcsecond resolution. 

This is the first in a series of papers describing high-resolution imaging with LOFAR. In this paper, we will cover the overall calibration strategy that will prepare the data for high-resolution imaging. Paper II is a companion paper, which covers the complete Long Baseline Calibrator Survey. Future papers will cover wide-area postage-stamp surveying by post-processing data from the LOFAR Two-metre Sky Survey \citep[LoTSS;][]{shimwell_lofar_2019} and a strategy for making a single wide-field image of the entire field of view at once. The ultimate goal is wide-field imaging of every pointing, which will open up an incredible discovery space, but as this is extremely computationally expensive, we can make great inroads in the meantime by postage-stamp imaging of individual sources. Both of these strategies use the pipeline described here as a basis for preparing the data for their more advanced calibration and imaging. 

The pipeline can also be used on its own to generate appropriately calibrated data for an observation of an individual target near or at the phase centre for PI-led projects. This pipeline is publicly available on Github\footnote{\href{https://github.com/lmorabit/lofar-vlbi}{https://github.com/lmorabit/lofar-vlbi}} with complete documentation on how to use it. It has been designed to enable a non-expert LOFAR user to produce sub-arcsecond images using the international LOFAR array for user-defined science targets in the field of view. The pipeline is designed to operate on both HBA and LBA data, although here we only test data from the HBA.

The paper is organised as follows. In Section~\ref{sec:observations} we describe the typical observational considerations (including the observation used in this paper), Radio Observatory pre-processing, flux density calibrator pre-processing, and pre-processing of the Dutch array for the target observation. Section~\ref{sec:overview} starts with a flowchart of the calibration strategy, and describes the unique considerations for high-resolution imaging with LOFAR. The high-resolution imaging pipeline is described in Section~\ref{sec:pipeline}, with post-pipeline steps described in Section~\ref{sec:postpipeline}. Finally, results are presented in Section~\ref{sec:results} with a summary and future work given in Section~\ref{sec:conclusions}.

\section{Observations and pre-processing}
\label{sec:observations}
To showcase this pipeline we selected P205+55, a typical field from the LOFAR Two-metre Sky Survey \citep[LoTSS;][]{shimwell_lofar_2019}. The data presented here are a re-observation of this field through a commissioning proposal to provide high-resolution models for standard flux density calibrators (PI: F. Sweijen). We used the standard LoTSS observational setup, which is to use a standard flux density calibrator bookended on either side of the 8-hour on-source time. For this observation, the standard flux density calibrator was 3C 147, which is compact on $\sim1$\sarc\ $\,$scales. The standard flux density calibrators typically used by LOFAR, in order of most compact to least compact, are: 3C 48, 3C 147, 3C 295, 3C 196. Recently high-resolution models for 3C 295 and 3C 196\footnote{3C 295 model courtesy of F. Sweijen and 3C 196 model courtesy of A. Offringa.} have become available. Although these are all known to provide good bandpass solutions when appropriate models are provided, it is preferred to use the most compact calibrator available for the HBA to get the best solutions. When moving to the LBA, the user should keep in mind  that 3C 48, 3C 147, and 3C 295 all have low-frequency spectral turn-overs that will reduce the signal-to-noise at the low end of the frequency range. 

We employed the LoTSS observation strategy that is described in detail in \cite{shimwell_lofar_2017}, so we only summarise the salient points here. The observation bandwidth is 120 - 168 MHz, which avoids the worst radio frequency interference (RFI) contamination at the high end of the band and the poorest system equivalent flux density (SEFD) at the low end of the band. The data were recorded with 1 second sampling and in channels of 3.05 kHz width (64 channels per subband, 244 subbands). The frequency resolution allows for better RFI excision before averaging the data, which the Radio Observatory performed as part of the data-preprocessing using the \textsc{AOflagger} \citep{offringa_aoflagger_2010}. The data were then averaged to 12.205 kHz per channel (16 channels per subband) before being archived in the LOFAR Long Term Archive (LTA). 

\begin{figure}
\begin{center}
\includegraphics[width=0.5\textwidth]{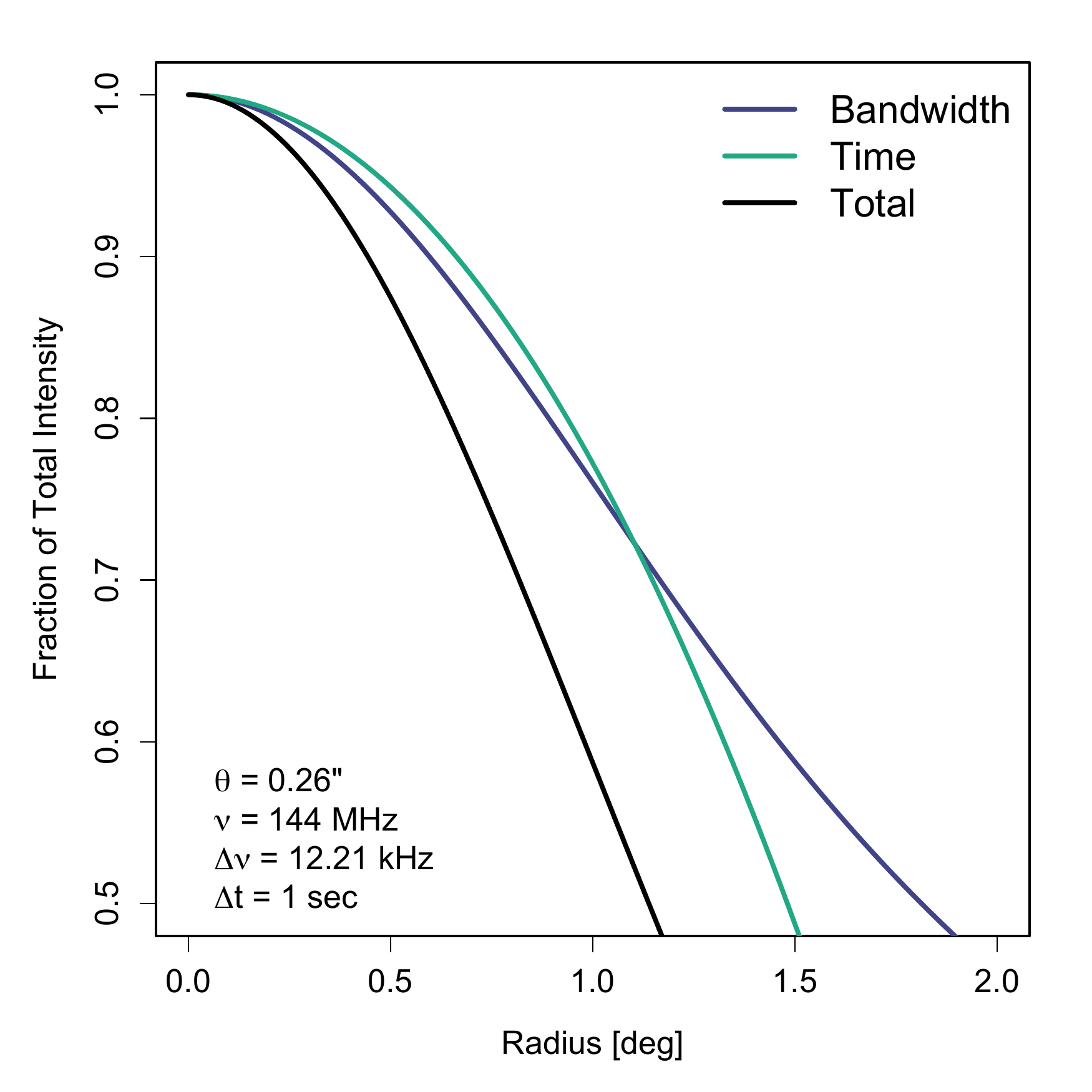}
\caption{\label{fig:intensitylosses} Fraction of initial intensity for a point source as a function of radius from the pointing centre, due to bandwidth and time smearing losses, as calculated from equations in \cite{bridle_bandwidth_1999}.}
\end{center}
\end{figure}

This averaging strategy mitigates bandwidth and time smearing to preserve the field of view as much as possible while still reducing the data volume. As a rule of thumb, a full data set averaged with these parameters will be 4$\,$TB with \textsc{dysco} compression \citep{offringa_compression_2016} and 16$\,$TB without compression. With time and frequency resolutions of 1 second and 12.205 kHz, the intensity losses from time and bandwidth smearing \citep[using equations 18-43 and 18-24 of ][]{bridle_bandwidth_1999} are roughly equal, as shown in Figure~\ref{fig:intensitylosses}. The bandwidth smearing is determined by the frequency resolution and is baseline dependent with the longest baseline being the most effected (for P205+55, this is Ireland to Poland, 1989 km). The combined intensity losses from bandwidth and time smearing are 20 percent at a radius of 0.65 degrees from the phase centre, and drop to 50 percent at 1.14 degrees. Although the bandwidth smearing can be mitigated by retaining the data at the typical 64 channels per subband, the field of view will still be limited by the time smearing, and this should be kept in mind when determining the observational strategy.

\begin{figure*}
\begin{center}
\includegraphics[width=\textwidth,clip,trim=0.5cm 0cm 0cm 0.5cm]{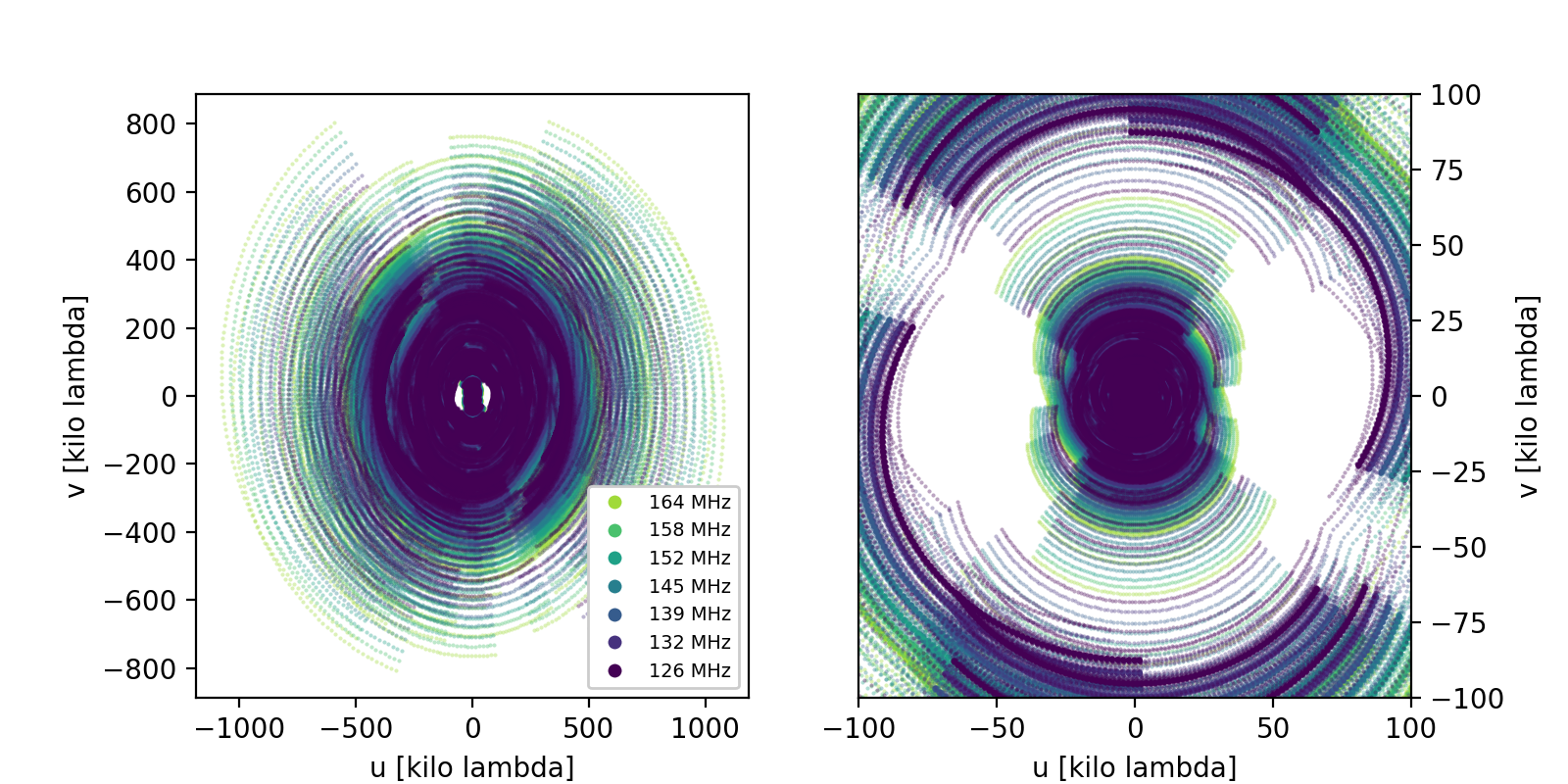}
\caption{\label{fig:uvcoverage}$u$-$v$ coverage of the observation. Plotted are six discrete frequencies and one sample every 5 minutes of the observation, including complex conjugates. On the left is the full $u$-$v$ coverage, while the right panel shows a zoom in of the inner 100 k$\lambda$ ($\sim$200 km). The gap here corresponds approximately to scales of 25\sarc\ to 12\sarc\ .}
\end{center}
\end{figure*}

The Dutch stations can be configured for different observational modes, while the international stations always use all 96 antenna elements (which provides increased sensitivity and decreased field of view when compared to stations in the Netherlands). All of the Dutch stations contain a total of 48 antenna tiles but in the core stations these are divided equally into two sub-stations. The data were taken with the \textsc{hba\_dual\_inner} mode, which uses all 24 tiles in each core sub-station as a separate antenna, while the remote stations use only their inner 24 (out of 48 total) antenna tiles. This yields a total of 75 stations for the observation. The observations are summarised in Table~\ref{tab:obs}. The $u$-$v$ coverage of this observation is presented in Figure~\ref{fig:uvcoverage}.

\begin{table}
\caption{\label{tab:obs}Observation parameters.}
\begin{tabular}{ll}
\hline \\[-6pt]
\multicolumn{2}{c}{General} \\ \hline \\[-8pt]
Date & 25 January 2019 \\
Mode &  \textsc{HBA\_DUAL\_INNER} \\
Core stations & 2$\times$24 \\
Remote stations & 14 \\
International stations & 13 (pre Latvian station) \\
$\Delta t$ & 1 sec \\
$\Delta \nu$ observed & 3.05 kHz \\
$\Delta \nu$ recorded & 12.205 kHz \\
Total bandwidth & 48 MHz \\ \hline \\[-6pt]
\multicolumn{2}{c}{Target} \\ \hline \\[-8pt]
P205+55 & 13:40:13.20 +54.53.53.52 \\
Observation ID & L693725 \\
On source time & 8 hours \\ 
Elevation & $>60$ degrees \\ \hline \\[-6pt]
\multicolumn{2}{c}{Calibrator} \\ \hline \\[-8pt]
3C 147 & 05:42:36.26 +49.51.07.08 \\
Observation ID & L693719 \\
On source time & 10 minutes \\ \hline 
\end{tabular}
\end{table}

The high resolution calibration strategy will be described in the next section, but it is designed to be compatible with the pipelines used by the LOFAR Two-metre Sky Survey (LoTSS). We summarise these pipelines and their outputs in terms of the international stations here. A summary of the computing resources necessary for this processing is given in Appendix~\ref{app:profile}. 

\subsection{Standard calibrator processing}
Finding the gains and setting the flux density scale using a standard calibrator can be accomplished using the \prefac \footnote{\url{https://github.com/lofar-astron/prefactor}} pipeline {\tt Pre-Facet-Calibrator.parset}. In this paper we have used commit 7e9103d of the master branch. This commit of the pipeline includes all international stations in the processing by default, which has negligible impact on the Dutch station solutions. However, this commit does not have the high-resolution models appropriate for 3C 196 and 3C 295\footnote{These have now been included in \prefac\ commit 79116d7.}; as our calibrator is 3C 147 this is not an issue for P205+55\footnote{This specific \prefac\ commit, which is known to be compatible with the \lofvlbi\ pipeline, has been forked at \href{https://github.com/lmorabit/prefactor}{https://github.com/lmorabit/prefactor} and the skymodels updated to include high-resolution models for all standard flux density calibrators.}. As the documentation for \prefac\ can be found at the Github repository, and a detailed description of the calibration strategy is given in \cite{de_gasperin_effect_2018}, here we simply summarise the relevant points. 

The calibrator pipeline by default first flags and averages the data. The calibration is then performed in incremental steps; first solving for the complex XX and YY gains and then extracting an effect from the solutions, before correcting for that effect and repeating the process using the corrected data. The following effects are found, in this order: polarisation alignment (correcting differences between XX and YY), rotation measure (from Faraday rotation), bandpass, clock, and total electron content (TEC) values. The direction-independent effects (polarisation alignment, bandpass, and clock) can then be transferred to the target data. It is important to correct the amplitude gains for the international stations as early as possible, so all stations are on the same relative scale, and the flux density scale is tied to a well-known calibrator source. Using standard flux density calibrators ensures that that the gains are found in a high signal-to-noise regime, which increases the quality of the solutions. 

\begin{figure*}
\begin{center}
\includegraphics[width=0.45\textwidth]{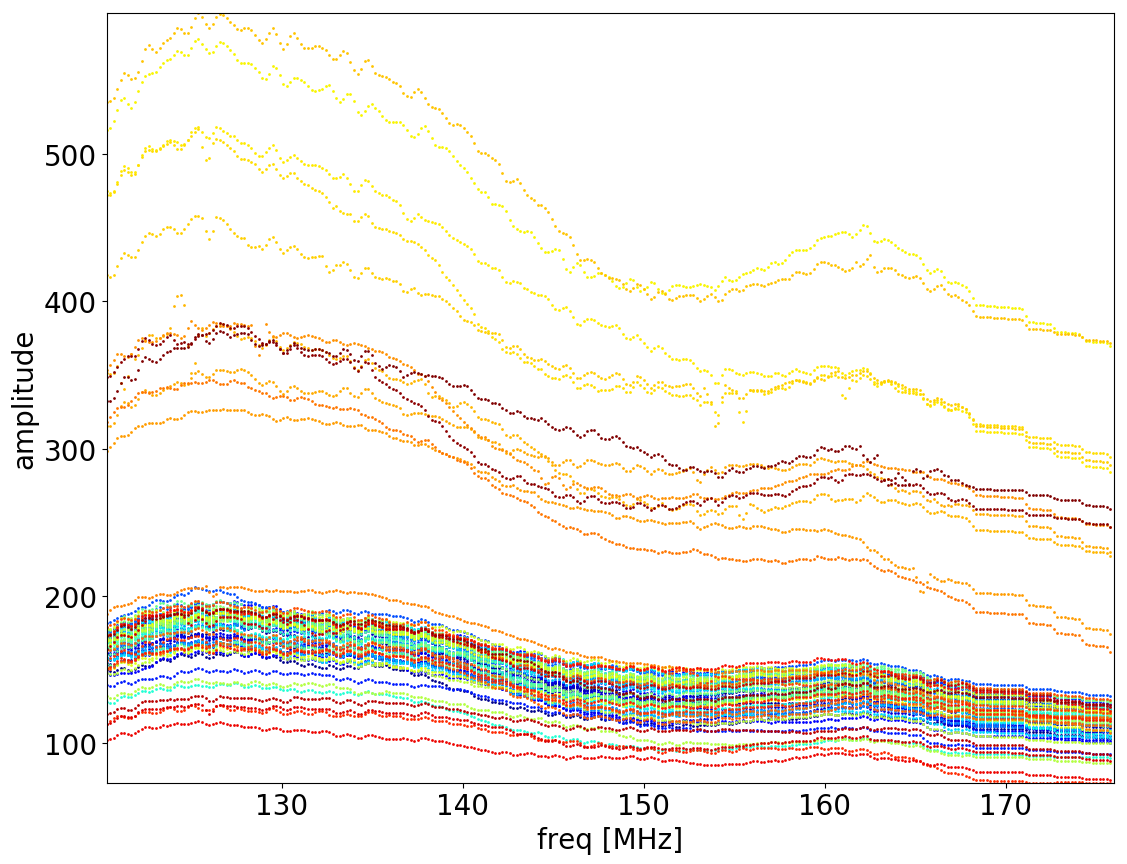} \hfill
\includegraphics[width=0.5\textwidth]{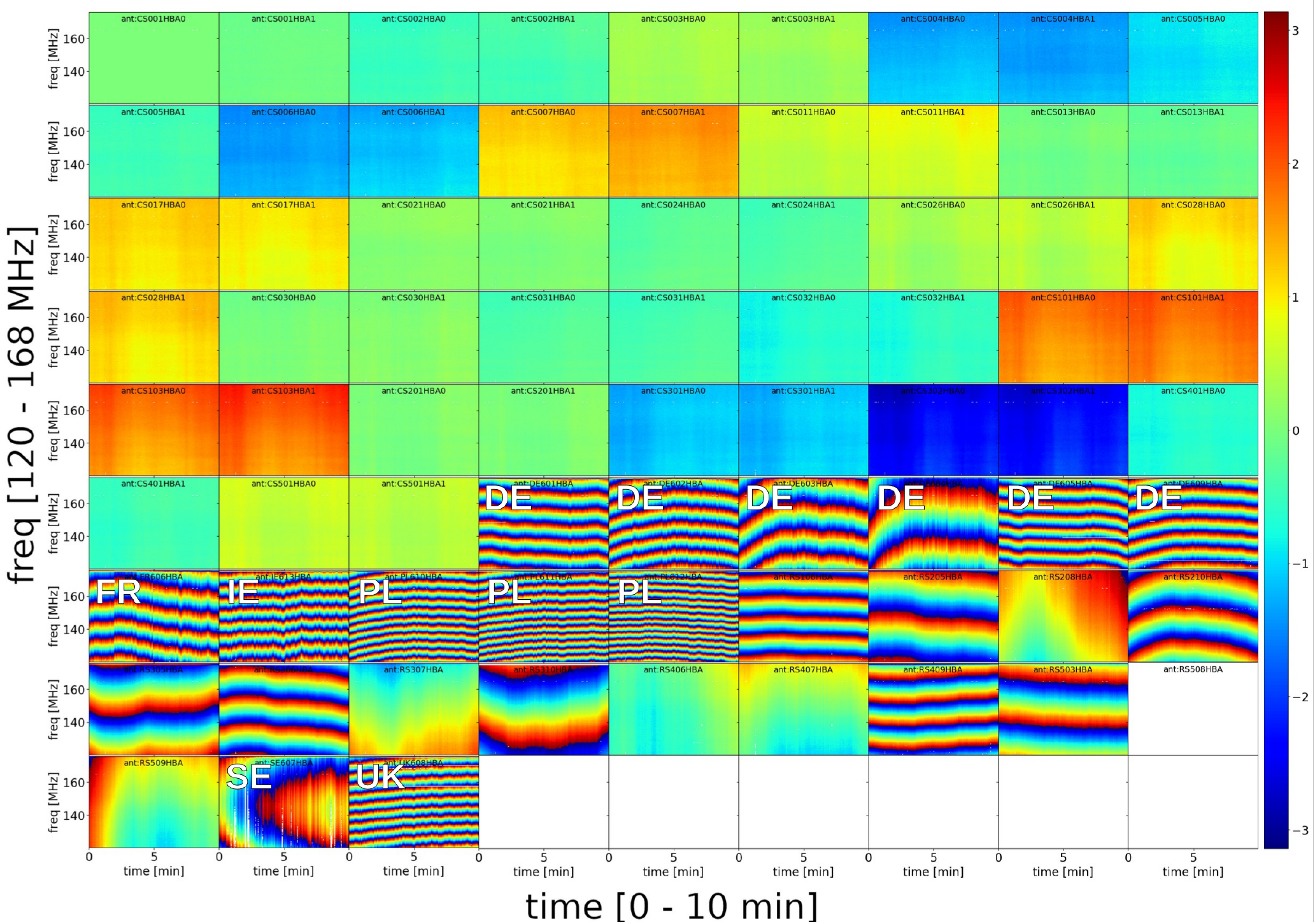}
\caption{\label{fig:calibrator}Calibrator solutions for 3C 147. These are two of the default plots generated by \prefac . On the left are bandpass solutions for the XX polarisation. The international stations are those with values of $\sim$300 or above, while the Dutch station bandpass values are clustered between 100 - 200. On the right are phase solutions for the XX polarisation, referenced to CS001HBA (top left corner). Each of the sub-panels in the right hand panel share the same axes: 0 - 10 minutes for the time axis, and 120 - 168 MHz for the frequency axis. It is clear that the core stations (CS) and remote stations (RS) have phases that are close to each other and vary slowly in frequency, while the international station phases (labelled with country codes DE, FR, SE, IR, PL, UK) show faster variation with frequency (in the vertical direction). The amplitude values are a correction factor that fixes the data scale to be in units of Janskys. The units of phase are radians. }
\end{center}
\end{figure*}

The international station phases will vary more quickly with frequency and time, and the bandpasses (which fix the data amplitude scale to be in units of Janskys) typically have values that are a factor of 2-3 higher than the Dutch stations. The reason for higher values on international compared to core and remote stations is because the international stations have more tiles and are more sensitive; the bandpass amplitude approximately reflects the station gain. It is therefore important to check the \prefac -generated inspection plots to ensure that default values for flagging or clipping have not removed international stations erroneously. It can be the case that an international station has corrupted data and should be removed, but this needs to be assessed carefully. Catastrophic failures are reported in the Radio Observatory observation logs. Figure~\ref{fig:calibrator} shows the bandpass and final phase solutions (from the {\tt ion} step) from 3C 147 for P205+55. 

\subsection{Direction-independent processing of Dutch stations}
Before starting the calibration of the international stations, we make use of \prefac\ solutions to correct the Dutch stations. The reason this is important is described in detail in Section~\ref{subsec:introcombination}. The pipeline for the target ({\tt Pre-Facet-Target.parset}) operates only on Dutch stations, which reduces the number of baselines (and therefore data volume) by about a factor of six.  The pipeline begins by removing the international stations, flagging the data, and removing bright off-axis sources via demixing if necessary. Demixing is a process where the data are phase-rotated to the nearby bright off-axis source, averaged down in frequency and time, calibrated against a model; these solutions are then used to subtract the source from the data \citep{van_der_tol_self-calibration_2007}. After this, the rotation measure for each station, including the international stations, in the direction of the target field is downloaded from Global Positioning System (GPS) data using the \textsc{RMextract} package \citep{mevius_rmextract_2018}. These data only have measurements at 15 minute intervals, but it does provide an improvement by correcting for bulk changes in the Total Electron Content (TEC) in preparation for solving for differential TEC at a later stage in the calibration process (Section~\ref{subsec:phaseref}). 

Corrections are applied to the target in the following order: polarisation alignment, bandpass, clock, beam, and rotation measure. The data are then averaged by a factor of four in both frequency and time. The contributions of bright off-axis sources that were not demixed are estimated and clipped from the data. Finally, the pipeline performs a phase-only calibration against a skymodel extracted from the TGSS-ADR1 \citep{intema_gmrt_2017}. This survey is at 150$\,$MHz and has a resolution of 25\sarc\ . 

The final steps of the target pipeline collect the calibrator and target solutions into a single file in Hierarchical data format version 5 \citep{hdf5}, which we call an h5parm. For the target, this comprises the rotation measure for all stations (including international) and the phase-only solutions for the Dutch stations. Up to this point, there is only one phase solution found for the entire field per band (of ten subbands each), providing a direction-independent correction. 

\subsection{Direction-dependent processing of Dutch stations }
\label{subsec:ddfpipeline}
The direction-independent calibrations for the target field can be improved on by correcting DDEs. Section 2.3 in \cite{shimwell_lofar_2019} gives a practical overview of the topic as it applies to LOFAR data. Essentially the pipeline makes use of Jones matrices \citep{hamaker_understanding_1996} to encapsulate time, frequency, antenna, and direction dependent effects. Using KillMS \citep{Tasse_applying_2014,smirnov_radio_2015} to calculate the Jones matrices and \textsc{ddfacet} \citep{tasse_faceting_2018} to apply them during imaging, several rounds of self-calibration are performed. We used the LoTSS-DR1 \ddfp\ setup\footnote{\href{https://github.com/mhardcastle/ddf-pipeline}{https://github.com/mhardcastle/ddf-pipeline}}, which performs three rounds of self-calibration, and includes bootstrapping the flux density scale and correcting the astrometry as described in \cite{shimwell_lofar_2019}. 

While this step is optional, it does provide the user several important data products: Jones matrices with direction-independent corrections to refine the core and remote station solutions; a pointing-specific catalogue that is used in case LoTSS has not yet covered this part of the sky; and  a science-quality image at 6\sarc\ \ resolution, which is useful for pre-identification of sources. The final two points will become less relevant as LoTSS continues, but for now any user who wishes to process a data set through the \lofvlbi\ pipeline will need either a LoTSS catalogue or the output of this step to provide flux density information for LBCS calibrators (described further in Section~\ref{subsec:catalogues}). The Jones matrices are also required for combined wide-field imaging. This will be covered in a later paper in this series. 

\section{Calibration strategy for LOFAR-VLBI}
\label{sec:overview}
In this section we first provide an overview of the calibration strategy and introduce two important concepts that have shaped the strategy. The specific steps the pipeline employs will be discussed in depth in the next section.

\begin{figure*}[h]
\begin{center}
\includegraphics[width=0.9\textwidth,clip,trim=0cm 0cm 0cm 0cm]{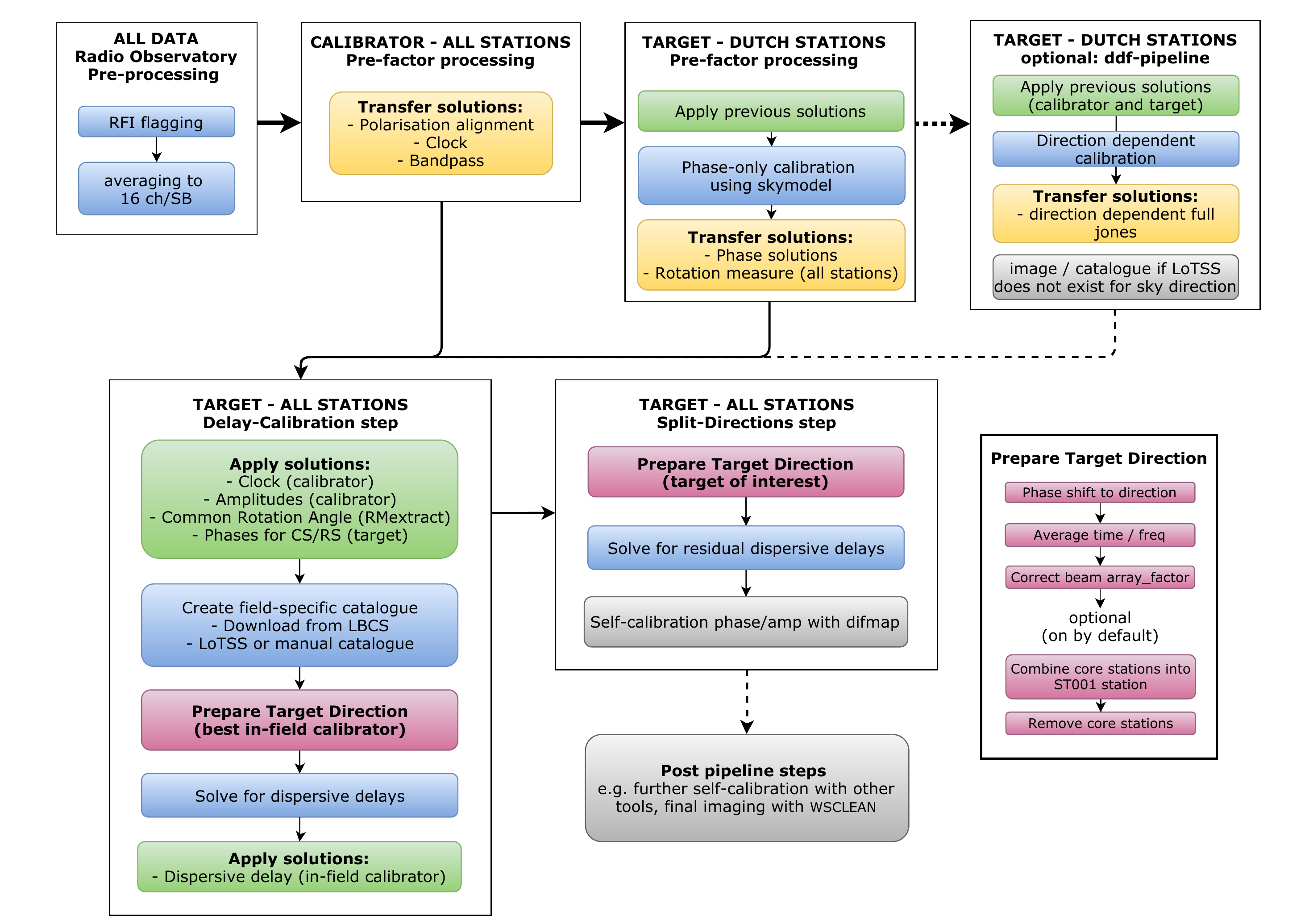}
\caption{\label{fig:blockdiagram} Block diagram overview of the calibration strategy. Each white box represents a different self-contained step in the process, starting with Radio Observatory Pre-processing, all the way to the Split-Directions step of the \lofvlbi\ pipeline. Data products from previous boxes are used in the next box. At the top of each box the type of data and stations used are indicated in capital letters.}
\end{center}
\end{figure*}

\subsection{Overview of strategy}
The \lofvlbi\ strategy, which is depicted in the block diagram in Figure~\ref{fig:blockdiagram}, builds on the pre-processing described in the previous section. This enables us to take advantage of calibration solutions that have already been produced in the case of post-processing LoTSS data, or preparation of the data to the community standard if processing PI data. These solutions are applied to data that include all international stations, providing a data set for which all stations have been corrected for polarisation alignment, clock, rotation measure, and bandpass. The Dutch stations have additionally been corrected for direction-independent phases (using \prefac ) and optionally direction-dependent solutions (\ddfp ). 

The \lofvlbi\ strategy continues by building information on sources within the field, cross-matching the 6\sarc\ catalogue (either from LoTSS or manually generated) and LBCS. The pipeline determines which calibrator is best to solve for direction-independent dispersive delays. From there, the pipeline solves directly for TEC on the best calibrator candidate, and applies the solutions to the full-resolution data. After this, a list of imaging directions is used to phase-rotate the data to a particular direction, and split out time and frequency averaged data sets for self-calibration. The end result is smaller, calibrated data sets and preliminary images in the directions provided to the pipeline. 

Although the International LOFAR Telescope resembles a connected-element interferometer in that data transport and correlation occurs in real time, the baseline length and independent frequency standards at the international stations necessitate a treatment that closely follows that used for Very Long Baseline Interferometry (VLBI) at centimetre and millimetre wavelengths. We developed this calibration strategy based on VLBI principles, with updates for LOFAR-specific challenges using native LOFAR software wherever possible. Before discussing the specific steps in the calibration strategy, we introduce two main concepts which have driven our strategy.

\subsection{Introduction: Delay calibration}
\label{subsec:introdelays}
Phase errors have a dependence on both frequency and time. By taking a first order expansion of the phase error, we can see this dependence explicitly:
\begin{equation}
\Delta \phi_{\nu,t} = \phi_0 + \left( \frac{d \phi}{d \nu }\Delta \nu + \frac{d \phi}{d t}\Delta t \right).
\end{equation}
The first term on the right hand side is a phase offset, while the second and third terms are known as the delay and rate terms, corresponding to how the phase errors change with frequency and time, respectively. Typically the quantity $\Delta \phi_{\nu,t}$ is solved for directly, for small time and bandwidth intervals, given high enough signal-to-noise in each solution interval. However, phases on longer baselines vary more rapidly than on short baselines. This variation can happen in both time and frequency, meaning that small solution intervals are necessary to track the changes. This drives the need for short solution intervals in time and small intervals of bandwidth. The combination of these requirements prohibits averaging in time and/or frequency to increase the signal-to-noise ratio (S/N). Furthermore, real astrophysical sources have spatial structure, and in the VLBI regime a large fraction of the emission can be resolved out, leaving only a small fraction of the signal in the compact, unresolved regions required for robust calibration. This drives the need to increase the S/N and hence the solution intervals. 

Using larger time and bandwidth intervals is possible if the delays and rates are solved for in addition to the phases; this technique is called `fringe-fitting' and is traditionally used in VLBI. Global fringe-fitting \citep{schwab_global_1983}, as implemented in \textsc{AIPS} \citep{greisen_aips_2003} uses all baselines simultaneously to find solutions for every antenna even if the S/N is too low on individual baselines to each antenna. However, the fringe-fitting in \textsc{AIPS} assumes no dependence between the delay and frequency, which is a good approximation at high frequencies ($\gtrsim 1\,$GHz) but not appropriate at low frequencies ($\lesssim 300\,$MHz). When this pipeline was already in an advanced stage, a fringe-fitting algorithm that accounts for dispersive (non-linear dependence between phase and frequency) delays was implemented in \textsc{CASA} \citep{van_bemmel_casa_2018}. By this point in time we had developed our calibration strategy based on dedicated LOFAR tools to perform robust calibration on LBCS calibrators without this, but it will be useful in the future for fainter science targets. 

The frequency solution interval, $\Delta \nu$, will depend on how quickly the phase changes with frequency. This is set by two dominant types of delays in the data: offsets in the clocks at different stations and dispersive delays from propagation of the radio waves through the ionosphere. The clock offsets produce non-dispersive delays, which means that the delay ($\tau_{cl} = d\phi/d\nu$) does not change with frequency. This is an instrumental effect, and will be direction independent; that is, the clock offsets are the same no matter which direction the telescope is pointing. Ionospheric delays are dispersive, with a first-order $\tau_{ion} = d\phi/d\nu \propto \nu^{-2}$  dependence, and will vary based on the path or propagation through the ionosphere. The area of sky over which the phases are coherent, and therefore the delays can be tracked, varies with ionospheric conditions. In some cases the ionospheric delays are similar up to a couple degrees, while in other cases even transferring solutions over half a degree might not be possible. The delays from clock offsets and propagation through the ionosphere are additive, although the ionospheric delays dominate, and their combined effect is imprinted in the data. 

The time solution interval, $\Delta t$, is generally set by the coherence time, which is the length of time over which the phase is predictable. While this is related to the wavelength, and in theory should be large, the ionosphere can vary rapidly, which decreases the coherence time. Finding good solutions is  a balancing act between increasing $\Delta t$ to solve for reliable rates, and keeping the solution interval within the coherence time so the phase and rate solutions are not degraded by decorrelation. For LOFAR observations including the international stations, coherence times are typically $\sim$2 minutes in good ionospheric conditions, but can be as short as 10-15 seconds in bad ionospheric conditions. We have found that for LBCS type `P' calibrators (expected to be good calibrator sources for all international baselines), there is enough signal-to-noise to track the phases and delays for small time intervals (over which we approximate the rates to be constant), in agreement with LBCS (Jackson 2021). 

It is useful to correct all instrumental effects before attempting to solve for DDEs. Thus we would like to independently correct for the delays introduced by the clock offsets. This can be done using an algorithm developed by M. Mevius and implemented in the LOFAR Software SOlutions TOol \citep[LoSoTo;][]{de_gasperin_systematic_2019}. The algorithm uses the fact that the clock delays are non-dispersive, $d\phi/d\nu = \mathrm{constant}$, while the ionospheric delays are dispersive, $d\phi/d\nu\propto\nu^{-2}$, to separate the effects in solution space. This separation is accomplished by simultaneously fitting for these two different behaviours in phase solutions. \prefac\ performs this on the standard flux density calibrator, where the signal-to-noise is high. The clock offsets can then be transferred to the target data. Clock offsets for the international stations are typically $\sim 100 - 250$ nanoseconds. Because a single median clock offset per antenna is transferred from the calibrator to the target data, there can be minor residual delays from clock drifts. A delay of $\sim$20$\,$ns will result in the decorrelation of phases over the bandwidth of the observation ($\Delta\nu=48\,$MHz). The residual delays from clock drifts are typically $\lesssim 10\,$ns. However, the residual delays for sources in the field are driven by the ionosphere, and we see that sources $\sim$degrees away from each other may differ by $\gtrsim 20\,$ns. Solving for the residual non-dispersive delays in different directions may be necessary. This is not yet implemented in the pipeline, but is an area for future development. 

Determining and correcting the dispersive delays caused by the ionosphere (measured as TEC) is the main challenge for our calibration strategy. Because this effect can vary significantly between two points on the sky, it is necessary to estimate the ionospheric delays from a calibrator source near the scientific target. The sky density of suitable calibrator sources is not high, and sometimes we must resort to using fainter sources for phase calibration. This places us squarely in the regime of needing to use a technique such as fringe-fitting to increase the signal-to-noise. Previous LOFAR-VLBI publications have made use of fringe-fitting by dividing their bandwidth up into small enough chunks so the approximation that the delays are linear in frequency is appropriate. However, by solving directly for the dispersive delays due to the ionosphere, we can first remove the frequency dependence in the phases and subsequently use the entire bandwidth to perform phase calibration. We do this by using the TEC solving routine incorporated in the default post processing pipeline \citep[\ndppp ;][]{van_diepen_dppp_2018}. This fits the phases directly for $\phi \propto \nu^{-1}$ behaviour, and the result is expressed in terms of TEC. The conversion between TEC and phase is given by:
\begin{equation}
\label{eqn:tec}
\phi =  \phi_0 - \frac{c r_e}{4 \pi}\frac{\textrm{TEC}}{\nu},
\end{equation}
where $c$ is the speed of light, $r_e$ is the electron radius, $\nu$ is frequency, and TEC is in units of 10$^{16}\,$electrons$\,$m$^{-2}$.  Equation~\ref{eqn:tec} shows that a single value of TEC describes the phase behaviour at all frequencies. 

We compared the delays obtained as TEC values with other methods of solving for delays, on an in-field calibrator selected from LBCS, and using an appropriate bandwidth for each method. These included: \textit{(i)} fringe-fitting in \textsc{AIPS}; \textit{(ii)} using {\tt gaincal} in \textsc{casa} (mode `K'); \textit{(iii)} using the clock/TEC fitting algorithm implemented in LoSoTo; and \textit{(iv)} solving directly for TEC using \ndppp . We found that all of these methods produced very similar results, with only small quantitative differences arising from differences in the data preparation and/or algorithms used to find the delays. However, although the ideal situation would be to solve for clock (here we mean the clock drift as the offset is already corrected) and TEC values separately, the clock/TEC separation does not work in all cases on in-field calibrators. The pipeline therefore currently only solves for the dominant effect, which is the dispersive delay (TEC; option \textit{iv} above). Using the dedicated LOFAR software tool to do this has the advantage that we do not have to process data through software that is not optimised for LOFAR data. By this, we refer both to the fact that LOFAR observations produce large data volumes, which AIPS and CASA are not necessarily equipped to handle efficiently, and the fact that conversion between data formats and use in other software runs the risk of meta-data being missing or not properly handled. 

\subsection{Introduction: Combination of core stations}
\label{subsec:introcombination}
Another important technique we use in high-resolution imaging with LOFAR is combining the core stations into a `super' station. This in effect provides a super station (ST001) in the centre of the array that is $N_{\textrm{core}}$ times more sensitive, providing an anchor for calibrating the international stations. We have found that typical LOFAR calibration strategies are unable to successfully find good initial gain solutions in all cases for even the nearest German stations without using ST001. This technique is more important to use towards the beginning of the calibration strategy, for example for the delay calibration, when the phases have yet to be corrected.  There are 24 core stations in the Netherlands, all within a 4$\,$km radius, and in the \textsc{hba\_dual\_inner} mode each station is split into two independent sub-stations. This very compact core means that baselines from an international station to any core station will provide approximately the same information, that is, the baseline visibilities will be very closely spaced in the $u$-$v$ plane. We can therefore combine the core stations by finding the weighted average of all visibilities from a non-core station to all core stations. The weighted $u,v,w$ coordinates of the new visibilities are also calculated. The combination has to be done after any phase-rotating to different directions in the field of view (see Section~\ref{subsec:delaycalibration}). 

To combine the core stations coherently, they must be corrected for instrumental and phase errors first. This is done by transferring the \prefac\ solutions for the Dutch array to the data set with the international stations. Once ST001 is created, we remove all core stations from the data. This reduces the data volume by $\sim 80$ percent, which speeds up the rest of the data processing. Removing core stations also effectively removes the flux on shorter baselines, which can otherwise cause deconvolution problems when imaging (although this flux could also be suppressed with an inner $u$-$v$ cut). Combining the core stations together produces an effectively larger station with a substantially reduced field of view, making any self-calibration less sensitive to other sources in the field. 

\begin{figure}
\centering
\includegraphics[width=0.5\textwidth]{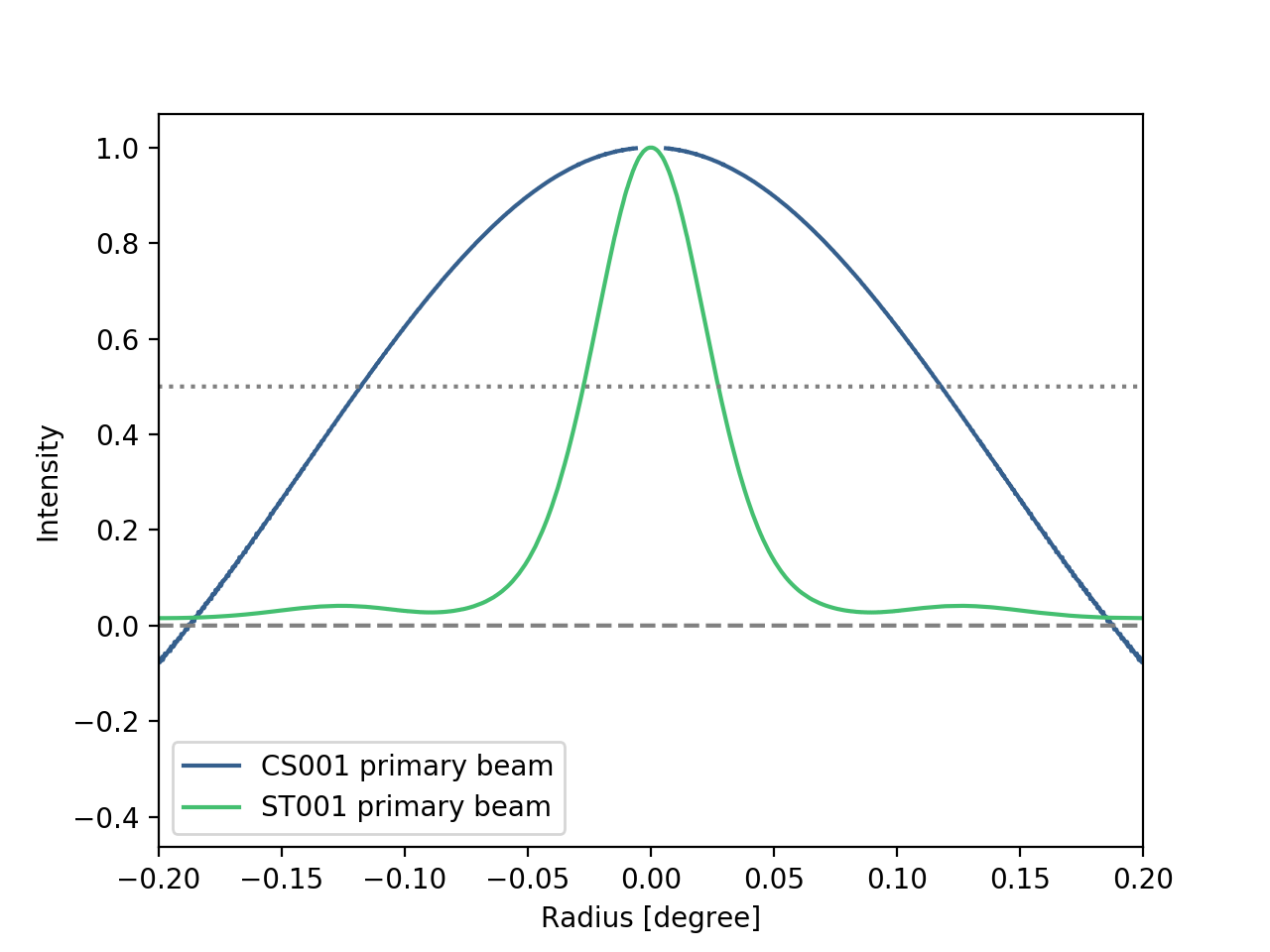}
\caption{\label{fig:beam}Comparison of the ST001 primary beam with a single core station primary beam. The beam profiles were created by imaging with natural weighting using all the core stations (ST001) and only the two-substations in a single core station (CS001). Only 10 iterations of \wsclean\ were run to provide an estimate of the PSF in both cases. The FWHM of each primary beam is: 100\sarc\ (ST001) and 425\sarc\ (CS001). Final images using the entire array will also be impacted by bandwidth and time smearing, which is a multiplicative factor. The dashed and dotted horizontal lines are drawn at zero and 0.5, respectively.}
\end{figure}

There is a drawback to combining the core stations; the combination introduces a radially varying decoherence on baselines containing ST001 \citep{bonnassieux_decoherence_2020}. This degrades the image fidelity, although for small (tens of arcsec) images the effect is negligible. 

\section{\lofvlbi\ pipeline}
\label{sec:pipeline}
In this section we describe each step of the pipeline in detail.  In this paper we use the V3.0.0 release of the \lofvlbi\ pipeline, which was run using a Singularity image. Singularity is a software container, which allows users to access a virtualised operating system that has all necessary software and dependencies in it. Information on software requirements, including how to access the specific singularity image we use\footnote{The singularity image can be found at \url{https://zenodo.org/record/4436416}.}, can be found in the documentation at \url{https://lofar-vlbi.readthedocs.io/}. 
Appendix~\ref{app:profile} provides basic pipeline profiling. 

\subsection{Catalogue generation}
\label{subsec:catalogues}
The \lofvlbi\ pipeline begins by constructing a model of sources in the field using the best available information, to collect information on LBCS calibrators, identify bright sources that may need to be subtracted to obtain good solutions, and identify a catalogue of sources that can be used for later imaging steps. The field of view is limited by bandwidth and time smearing on the longest baseline, and intensity losses drop to 50 percent at a radius of 1.14 degrees. We therefore limit our search for information to within 2 degrees, to capture bright sources outside this smearing limit but which may impact the data. The pipeline will automatically query both the LBCS\footnote{\url{https://lofar-surveys.org/lbcs.html}} and LoTSS\footnote{\url{https://vo.astron.nl/hetdex/lotss-dr1/cone/scs.xml}} online catalogues, using a default search radius of 2 degrees from the pointing centre. Although intensity losses drop rapidly, there can be extremely bright off-axis sources (e.g. 3C sources) that may also be suitable calibrators. It is necessary to query both databases as the LBCS catalogue does not contain flux density information, although this can be estimated based on the signal-to-noise to provide an approximate flux density measurement. For areas of the sky that are not yet fully covered by LoTSS, the user can generate their own catalogue and point the pipeline to a local file, specified by the {\tt lotss\_skymodel} parameter in the pipeline. If a local file is specified and exists, the pipeline will use this instead of querying the LoTSS database. For this paper, there was incomplete coverage of the field in LoTSS DR1 and we generated our own catalogue from the upcoming LoTSS Data Release 2 (Shimwell et al., in preparation). 

\begin{figure}
\begin{center}
\includegraphics[width=\columnwidth]{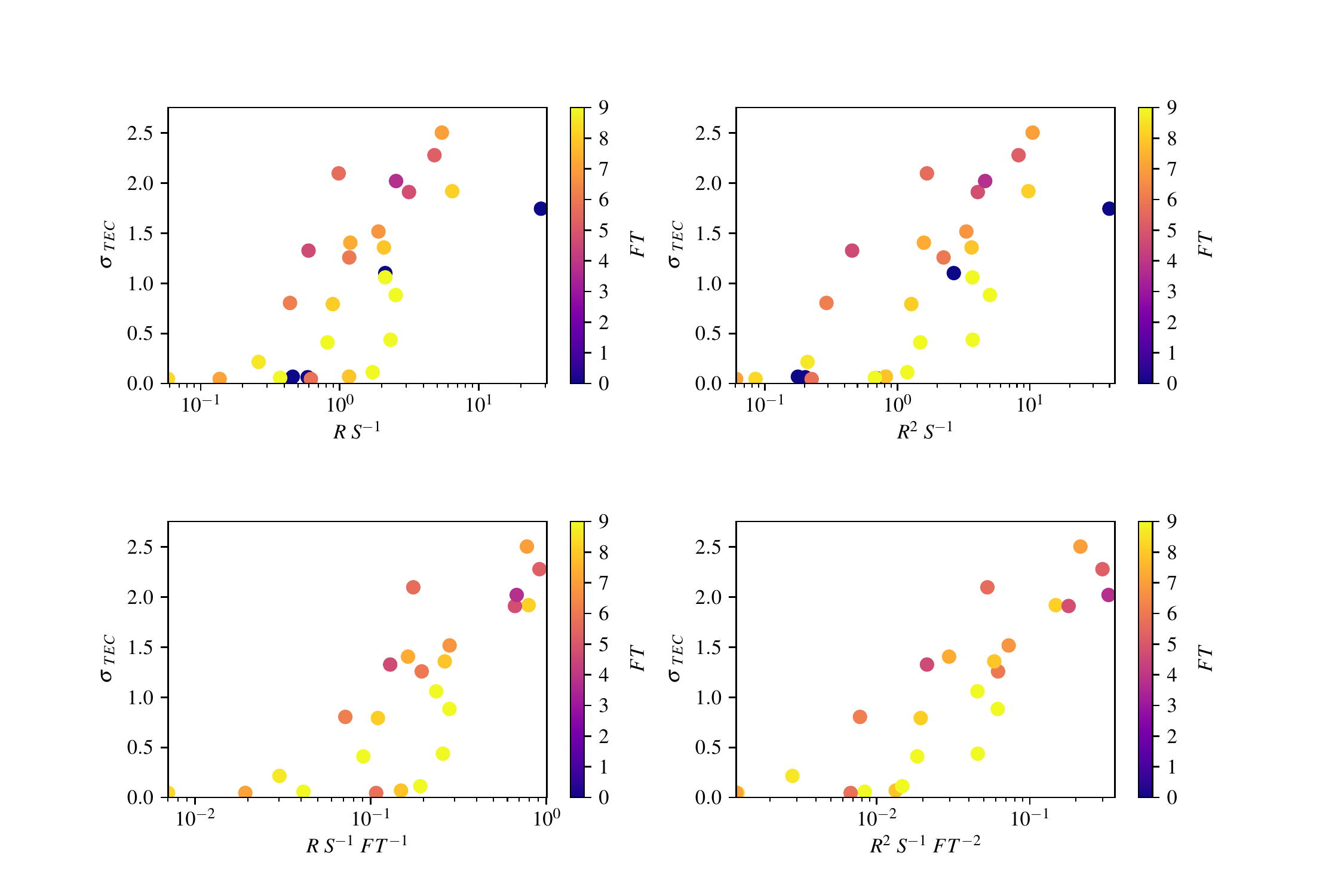}
\caption{\label{fig:tecscatter} Different combinations of radius ($R$), flux density ($S$), and an LBCS goodness statistic ($FT$). Each data point is a different LBCS calibrator from a combination of five different observations. The independent variables on each plot are the different combinations, with the scatter in the solutions plotted as the dependent variable. The strongest correlation, and therefore best predictor for low scatter in TEC solutions, is seen in the bottom right panel.}
\end{center}
\end{figure}

The pipeline cross-matches the LBCS and LoTSS (or user-generated) catalogues and writes out several key results. The LBCS sources are all considered to be potential in-field calibrators, from which the pipeline selects the best candidate to use for the delay calibration. To determine the selection algorithm for the best in-field calibrator, we treated all LBCS candidate calibrators in five different fields (observed with LOFAR HBA in an identical or very similar setup to P205+55) as the best in-field calibrator, and solved for the dispersive delays on each. This provides statistics for different distributions of sources in a field, different ionospheric conditions, and different declinations. We then examined different combinations of flux density, radius from phase centre, and different quality indicators from LBCS for the international stations. We plotted these combinations against a measure of the coherence of the dispersive delays, which we quantified as the scatter in TEC solutions. This was calculated as the median standard deviation of the solution values in a running window, using a local fit to the solutions as a function of time. Figure~\ref{fig:tecscatter} shows the different combinations we tested. The most robust predictor of low scatter in TEC solutions is
\begin{equation}
\label{eqn:2}
\frac{R^2}{S\times FT^2}, 
\end{equation} 
where $R$ is radius from phase centre, $S$ is total flux density, and $FT$ is a measure of signal-to-noise in the Fourier transformed data from LBCS. The S/N is assessed per international station as a value between 0 to 9 (inclusive). We added up the values for each station and normalised by the number of stations as not all international stations participate in every observation. By finding the source with the minimum total value of Equation~\ref{eqn:2}, we maximise the likelihood that source is the best in-field calibrator for a particular pointing. In all five fields, this predicted the candidate delay calibrator that would have been chosen from visual inspection of the solutions. 

Aside from selecting the best calibrator candidate, the pipeline will automatically prepare two other catalogues. The first is a list of all LBCS calibrators and bright sources (default $\geq 5\,$Jy but this is adjustable in the pipeline) that may need to be subtracted from the data to improve image fidelity. The second is an imaging catalogue, which contains a flux density sorted list of sources in the field. This can either be the source(s) to be directly imaged, or a list of sources that can be used to build up phase and TEC solutions in different directions across the field of view. The second case will be covered in a future paper in this series. If the user has specific targets to image, this final imaging catalogue can be provided manually instead, and specified in the pipeline as the {\tt imaging\_cat}. The pipeline will simply skip generating this (or any other) catalogue if it already exists. 

\subsection{Data preparation}
At the beginning of the {\tt Delay-Calibration} pipeline step, the pipeline prepares the data in the same manner as \prefac\ works on the Dutch stations, but including the international stations. This starts with applying the direction-independent \prefac\ solutions, both from the calibrator and the target. Next the pipeline performs clipping of data from interfering bright off-axis sources (although this step is lengthy and turned off by default; see pipeline profiling in Appendix~\ref{app:profile}), and then concatenates the data into bands of ten subbands ($\Delta\nu=1.95\,$MHz). We select the option to apply the \ddfp\ solutions, and then run RFI flagging using the \textsc{AOflagger} implementation in \ndppp . The end result of the data preparation are bands with $\Delta\nu=1.95\,$MHz that have corrected, flagged data in the {\tt DATA\_DI\_CORRECTED} ({\tt DATA}) column, if the \ddfp\ solutions were (not) used. These bands are used as the basis for the rest of the pipeline. 

\subsection{Delay calibration}
\label{subsec:delaycalibration}
The \lofvlbi\ pipeline next processes the best candidate for in-field delay calibration as identified in Section~\ref{subsec:catalogues}. We will refer to this in-field calibrator as the `delay calibrator'. Reading the information from the catalogue generation step, the pipeline creates a new measurement set that has been phase-shifted to the direction of the delay calibrator, averaging to two channels per subband and eight seconds. This averaging helps limit the field of view, which reduces contributions from other sources nearby. The core stations are then combined into ST001 and afterwards discarded to reduce the data volume. A beam correction (array factor) is performed to account for the time and frequency varying beam of each station that results when combining the dipoles / tiles together (for the HBA, this includes modelling the effect of the analogue tile beam former).

The dispersive delay calibration is performed in \ndppp\ using a point source model and the `tecphase' caltype implemented in the {\tt gaincal} step. We use a solution interval of $\Delta t=16\,$s with channels grouped into steps of $\Delta \nu=195\,$kHz (a single subband). Testing found that a minimum $u$-$v$ limit of 50k$\lambda$ (corresponding to $\sim$100 km) was important to reduce the number of outliers in the solutions, and had more success in producing coherent, high S/N solutions for international stations that were otherwise extremely noisy. This $u$-$v$ limit removes the contribution from the shortest baselines, suppressing the contribution of any nearby sources. 

The resulting solutions contain information on the difference in TEC amongst the stations, which we call differential total electron content (dTEC), and a single phase offset. We found that the phase offset was required to avoid introducing jumps in the dTEC solutions. The dTEC describes the phase behaviour as a function of frequency, and therefore a single frequency-independent dTEC value is used for phase corrections at all frequencies. Solving for dTEC requires fitting a curve to the phases along the frequency axis, and using the full bandwidth achieves the best performance, although only a minimum of 10$\,$MHz (about 30 subbands) is required to fit for dTEC. Here we use the full bandwith over 120 - 168$\,$MHz. Small gaps in frequencies due to failed or missing subbands do not impact the overall quality of the solutions. 

\begin{figure}
\begin{center}
\includegraphics[width=\columnwidth]{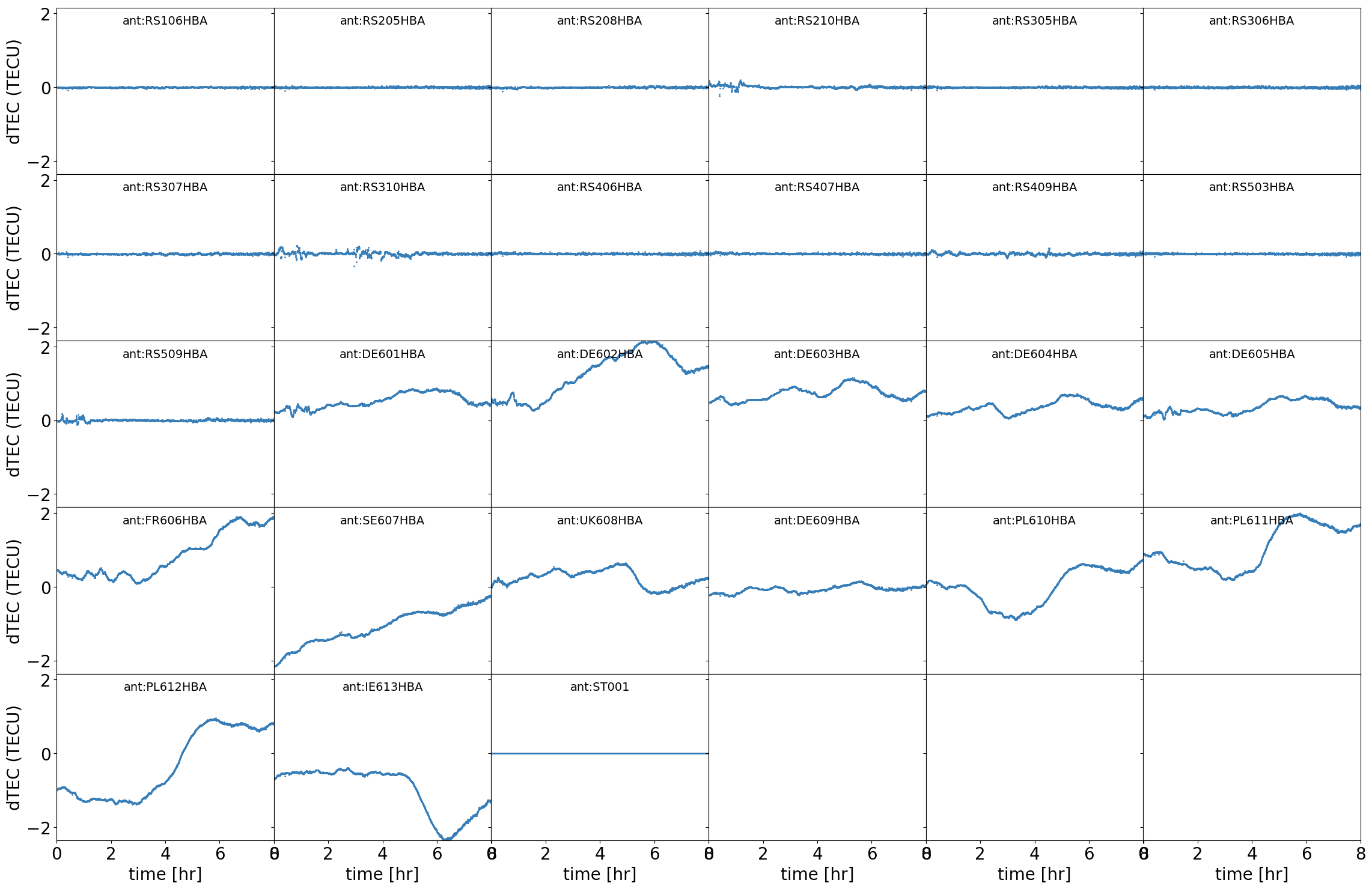}
\caption{\label{fig:delays} Dispersive delay solutions for the best in-field calibrator, J1337+5501. This is plotted in terms of differential TEC, which has units of 10$^{-16}$ electrons m$^{-2}$. The solutions are referenced to ST001, which is the final station in the plot. The remote stations (top two rows and first station on the third row) show relatively flat dTEC solution as these have already been corrected by \textsc{prefactor} .  }
\end{center}
\end{figure}

The best candidate for in-field calibration is likely to be bright, close to the phase centre, and/or compact. Reassuringly, we found that for these types of sources, using a point source model in the calibration returned a complex source structure after calibration and imaging, which converged during self-calibration. Several sources in the field were checked by hand to validate these results. Specifically in P205+55, the delay calibrator is close ($\lesssim 0.5\,$deg) to a strong ($3.3\,$Jy) source, but dTEC solutions did not improve after \textit{(i)} severely averaging to limit the field of view (averaged in time and frequency to produce 98 percent total intensity losses at a radius of 180 arcsec), or \textit{(ii)} subtracting the source. The pipeline therefore takes the simplest approach, using a point source model for the dispersive delay calibration without first subtracting nearby sources or averaging down to limit the field of view. We expect this to work in most cases, but we urge the user to inspect the dTEC solutions and manually select a different calibrator if necessary (as explained in the online documentation).
For a typical observation, the magnitude of dTEC on international stations can range from $\sim 0.1$ to 4 TEC units. For this observation, we see a span of 4 TEC units, from -2 to 2 TEC units (see Figure~\ref{fig:delays}). 
The remote stations show dTEC values around zero, which is expected as the \prefac\ corrections for dTEC have already been applied. 

Solving directly for dTEC yielded more stable solutions than first solving for phases and fitting for clock/TEC in solution space using LoSoTo. We attribute this both to working directly on noisy phases in individual channels in the visibility data, rather than working on per-channel solutions, which can be noisy, in solution space. Effectively this means we can use all of the bandwidth to find directly the dTEC values, rather than solving for phases in small chunks of bandwidth and operating on the solutions. For very good LBCS calibrators near the phase centre of the observation, both methods would be suitable, but we chose to implement solving for dTEC directly into the pipeline as it will work in a larger number of cases. That means that the clock drifts are not accounted for; however from testing we expect these to be typically $<10\,$ns. In the future, we plan to improve the pipeline by introducing newer methods that are able to account for both dispersive and non-dispersive delays.

The dTEC solutions are applied to the full resolution, un-phase-shifted data to which the \prefac\ solutions were already applied. The corrected data are written to the {\tt CORRECTED\_DATA} column, which is now ready for imaging and self-calibration of the science target.

\subsection{Self-calibration and imaging of directions of interest}
\label{subsec:selfcal}
At this point we can start imaging in any directions of interest (DOI) using the {\tt Split-Directions} step of the pipeline. The {\tt CORRECTED\_DATA} in the full-resolution, un-shifted data have been corrected for direction independent effects from \prefac\ and the dispersive delay from the in-field delay calibrator, as described in the previous section. In principle, the corrected data can be phase-shifted to the DOI, averaged down to reduce the field of view, corrected for the primary beam array factor, and then self-calibrated using whatever tool the user desires. However, while the delay correction removes the bulk of the dispersive delay, this can vary across the field and we recommend performing an initial TEC calibration in the DOI to solve for any residual dTEC values before moving on to standard self-calibration. The rest of this section describes the pipeline implementation of self-calibration, but the user can simply stop the pipeline after the residual delay calibration step, and perform their own self-calibration. We allow the self-calibration in our pipeline run, but also self-calibrate further as described in the next section. 

The pipeline reads in a user-supplied catalogue of DOI(s), which are treated independently, with the same steps carried out for each in parallel. To prepare the data in a DOI, the pipeline splits out a new measurement set, where the core stations are combined into ST001 (see Section~\ref{subsec:introcombination}), then flagged and removed from the data set. The data are also averaged to a frequency and time resolution of 97.64$\,$kHz and 8 seconds, respectively. The beam correction for the array factor is performed. The pipeline first runs these operations per band of 10 subbands, then combines all bands together into a single measurement set. It is worth noting that although combining the core stations into a super-station can lead to decorrelation in the final image, as described in \cite{bonnassieux_decoherence_2020}, we do not expect this to have a significant impact as the effect is radially dependent and we phase-shifted so our target of interest is at the phase centre before combining the stations. 

Each DOI is corrected for the residual dispersive delay (dTEC), which is the difference in dTEC between the delay calibrator and the DOI. Generally this is close to zero, but as $\phi \propto \nu^{-1}$, even small amounts of residual dTEC can cause the phase to decohere across 48 MHz of bandwidth. We model the DOI as a point source and solve for the dTEC and phase offset in the same manner as for the dispersive delay calibrator. It will not always be possible to do this as fainter sources may not have enough signal-to-noise to produce good solutions. In this case, solutions from the nearest LBCS calibrator can be found and applied to the science target before imaging. This is not currently a step in the pipeline and will have to be completed as a post-pipeline step (see Section~\ref{sec:postpipeline}).

The residual delays are applied to the DOI and then self-calibration is performed using \difmap\ \citep{shepherd_difmap_1997}, called from a python wrapper. \difmap\ is a purpose-built program built specifically for self-calibrating VLBI data, which was originally named after the `difference mapping' technique it uses. During a standard run of the \textsc{clean} algorithm, a model of the source(s) is built up by subtracting components iteratively from the residual map. This continues until the user stops the algorithm or a pre-set threshold has reached. Modifications to the model that is built are not allowed during the process; the user must start again if they find that \textsc{clean} has, for example, included too many imaging artefacts in the model. Difference mapping, on the other hand, allows the model and/or the visibilities to be modified during the \textsc{clean} process, by updating the residual map. This is calculated using the 2D Fast Fourier Transform (FFT) of the difference between the model and observed visibilities. Thus artefacts can be removed from the model as \textsc{clean} progresses, whereas the standard implementations of \textsc{clean} must re-start the entire process. 

The \difmap\ software is very compact and efficient, and typically more than an order of magnitude faster than other available tools. The disadvantages are that it is relatively rigid, and concentrates on doing only a few things really well; it does not write out corrections (only corrected data); and it self-calibrates amplitudes and phases but not delays. However, the pipeline corrects for delays by solving for and applying dTEC before passing the data to \difmap , and we found that the self-calibration out-performs our previous version using \ndppp\ and \wsclean\ \citep{offringa_wsclean_2014,offringa_wsclean_2017} . The \difmap\ solutions are written out using a modified version of the {\tt CORPLT} task, which is included in the singularity image. 

The python wrapper around the \difmap\ self-calibration converts the measurement set to uvfits format using {\tt ms2uvfits} with {\tt writesyscal=F}. Any completely flagged channels of data are identified, as these will cause \difmap\ to crash if they are processed, and a list of good channels is stored. The python wrapper writes a \difmap\ script and then calls it to select the data (Stokes I) and perform the self-calibration. The following steps are performed. First, a point source model is used for an initial self-calibration. Then, three rounds of phase self-calibration are accomplished with different $u$-$v$ weighting schemes. The clean / self-calibration loops proceed until the peak residual drops below a set fraction of the rms noise. After this, an overall amplitude scaling is performed followed by amplitude self-calibration with steadily decreasing solution intervals, and intervening phase self-calibration if the signal-to-noise of the output drops. The final clean map is generated and saved to a \textsc{fits} file and an ASCII table of phase and amplitude corrections are written. 

We note that this self-calibration scheme is only likely to work for strong sources, and we implement extra post-pipeline steps described later in Section~\ref{sec:postpipeline}. 
Once these steps are complete, the \difmap\ solutions are read in and converted to diagonal solutions, assuming that the polarisations XX and YY have equal contributions to Stokes I. We account for the fact that the \difmap\ and \ndppp\ conventions for phases differ by a sign change. The solutions are written out into a single h5parm and applied to the data, which is only a few GB after averaging to 1 minute and 0.39 MHz time and frequency resolution. This final h5parm can be updated to create solution tables for the core stations and copy the ST001 phase and amplitude solutions to each of them, although we continue to use ST001 in the post-pipeline steps described in the next section.

\section{Post-pipeline steps}
\label{sec:postpipeline}

The post-pipeline steps depend on the science goals, the distribution of calibrator sources and science target(s) in the field, and the data quality. We are in the process of developing the tools that will handle a wide variety of cases, but these are not yet implemented in a user-friendly pipeline, and will be the topic of a future paper in this series. Here we briefly outline the different cases that will dictate the current post-pipeline steps, from the simplest scenario to the most complex. 

\textbf{Case 1.} The user has a single science target at the phase centre, and the delay calibrator is within $\sim$ 1 degree, with typical ionospheric conditions. In this case, all self-calibration solutions from the delay calibrator should be transferred to the science target, using the measurement set with the combined super-station. From there the user can self-calibrate the science target if required or desired.

\textbf{Case 2.} The user has a science target not at the phase center, with the delay calibrator more than $\sim$ 1 degree away, or closer but with more rapid ionospheric variation. In this case, the user can try incrementally building up self-calibration solutions by applying these initially from the delay calibrator to a suitable LBCS calibrator closer to the science target, finding the residual solutions on the new calibrator, and apply all of the incremental solutions to the science target before the final imaging and self-calibration. 

\textbf{Case 3.} There are multiple science targets distributed across the field of view. In this case, all suitable LBCS calibrators should be used to find dTEC and self-calibration solutions, providing directional information on the TEC, phase, and amplitude variations across the field of view. These can then be interpolated to the positions of the science targets and applied before the final imaging and self-calibration.

The success of any of these scenarios will depend on the position of sources in any individual field and how active the ionosphere was during the observation. In poor ionospheric conditions, it may not be possible to transfer solutions even over short distances. Developing a pipeline that recognises and handles the majority of observations is ongoing work. Here we aim to demonstrate \textbf{Case 2} and validate some of \textbf{Case 3}, by transferring the delay calibrator dTEC and self-calibration solutions to all LBCS calibrators within 1.5 degrees of the phase centre and having a combined FT Goodness value $>$4. The FT Goodness is an S/N statistic that is calculated per antenna, with a value from 0 to 9, where higher numbers indicate better S/N. We combine these together by taking the average of antennas that participated in the observation. We selected a cutoff of 4 by plotting the  calibrator selection statistic described in Section~\ref{subsec:catalogues} against the combined FT Goodness, and found that below below 4 this statistic began to rise steeply, while above 4 the statistic was fairly flat, indicating good candidate calibrators. There are six calibrators that fit these criteria in P205+55. This strategy would also be appropriate for the simpler scenario of \textbf{Case 1}. Following self-calibration on the LBCS calibrators, we will finally image several directions in the field as test science targets. To do this, we use tools that already exist in the pipeline (e.g. software and scripts called by the pipeline) and \wsclean\ for imaging. 

\subsection{Further self-calibration on LBCS sources}
\label{subsec:furtherselfcal}
\begin{figure*}
\centering
\includegraphics[width=\textwidth]{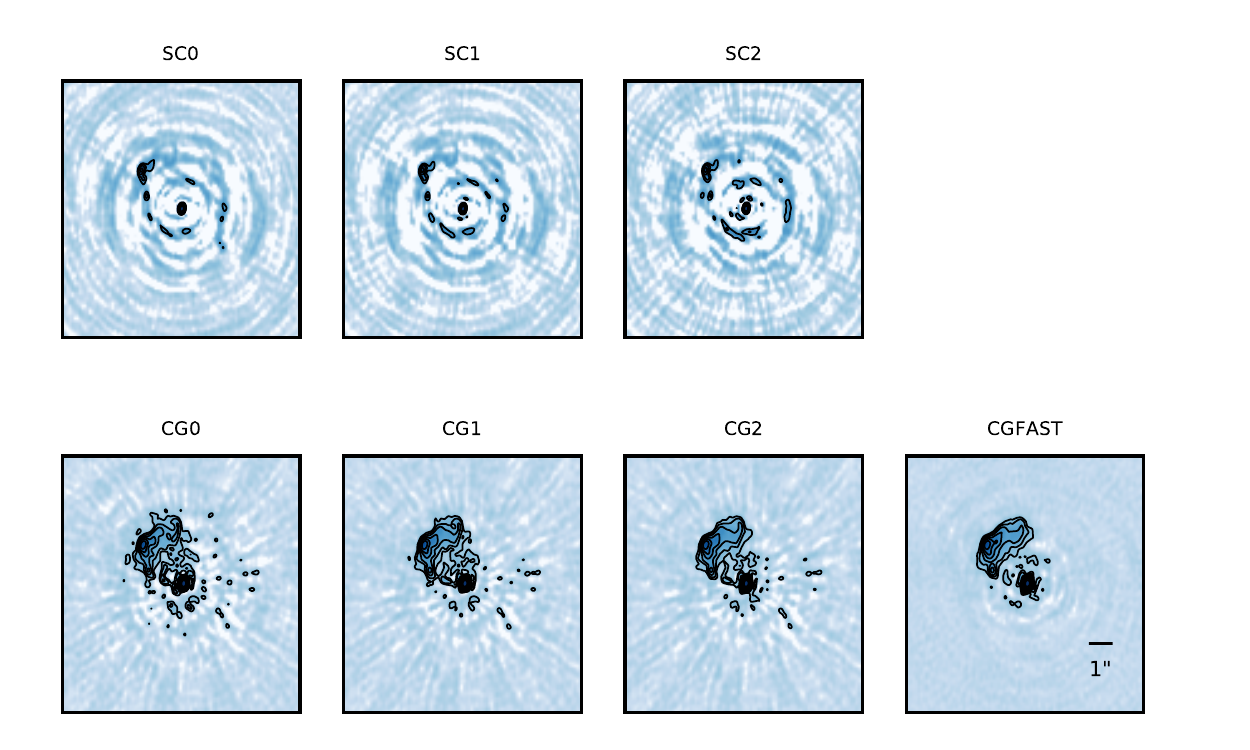}
\caption{\label{fig:dcselfcal}Self-calibration on the delay calibrator. The top row shows iterations of residual dTEC / \difmap\ self-calibration, starting with a point source and then using models generated from imaging the previous iteration. The bottom row shows complex gain self-calibration, with three iterations on solution intervals of 1 hour followed by a single complex gain iteration with a solution interval of 5 minutes. All plots are on the same colourscale with a log stretch to enhance the noise properties. The contours are set per row, starting at 10$\sigma$ and increasing by $\sqrt{2}$, based on the final image in the row.  }
\end{figure*}

The pipeline provides an initial data set in a direction of interest that has the dispersive delay corrected from the delay calibrator, with the solutions available from the final pipeline steps from an initial solve for residual dTEC and subsequent self-calibration with \difmap\ using a point source model. To improve these corrections, we can self-calibrate further to improve the solutions. In this subsection, we demonstrate our further self-calibration steps on the delay calibrator in P205+55, which is ILTJ133749.682+550102.754. The imaging from each step is shown in Figure~\ref{fig:dcselfcal}.  All operations are performed on the measurement set with the super-station, for two reasons. First, it is faster as the data volume is 14 percent that of the same measurement set with the core stations. Second, baselines including the super-station have increased sensitivity and help anchor the calibration of the international stations (see Section~\ref{subsec:introcombination}). 

The first step is to image the source to make a more accurate model. Applying the initial point-source model dTEC and self-calibration solutions, we image the delay calibrator and use \textsc{PyBDSF} \citep{mohan_pybdsf_2015} to extract a model and write it out in a format that is understood by \ndppp . Following this, we iterate the pipeline steps again, but using the source model generated from the source image. This consists of solving for the residual dTEC, which is applied. The corrected data are passed to the \difmap\ self-cal routine, which is run with the option to generate a starting model with an initial clean before self-calibration. Finally, we image the source with both the residual dTEC and \difmap\ self-calibration solutions applied. This entire process is repeated twice, for a total of three rounds of dTEC/\difmap\ self-calibration: once with a point source model and twice using a model generated from imaging the results of the previous iteration. Testing showed that a solution interval of $\Delta t=8\,$s yielded the best image fidelity at the end of the entire self-calibration, and also when transferring to other sources, although testing in other fields has found better results using longer solution intervals, related either to the ionospheric conditions and/or the calibrator S/N, so it is important to check this for each observation. 

\begin{figure}
\begin{center}
\includegraphics[width=0.5\textwidth]{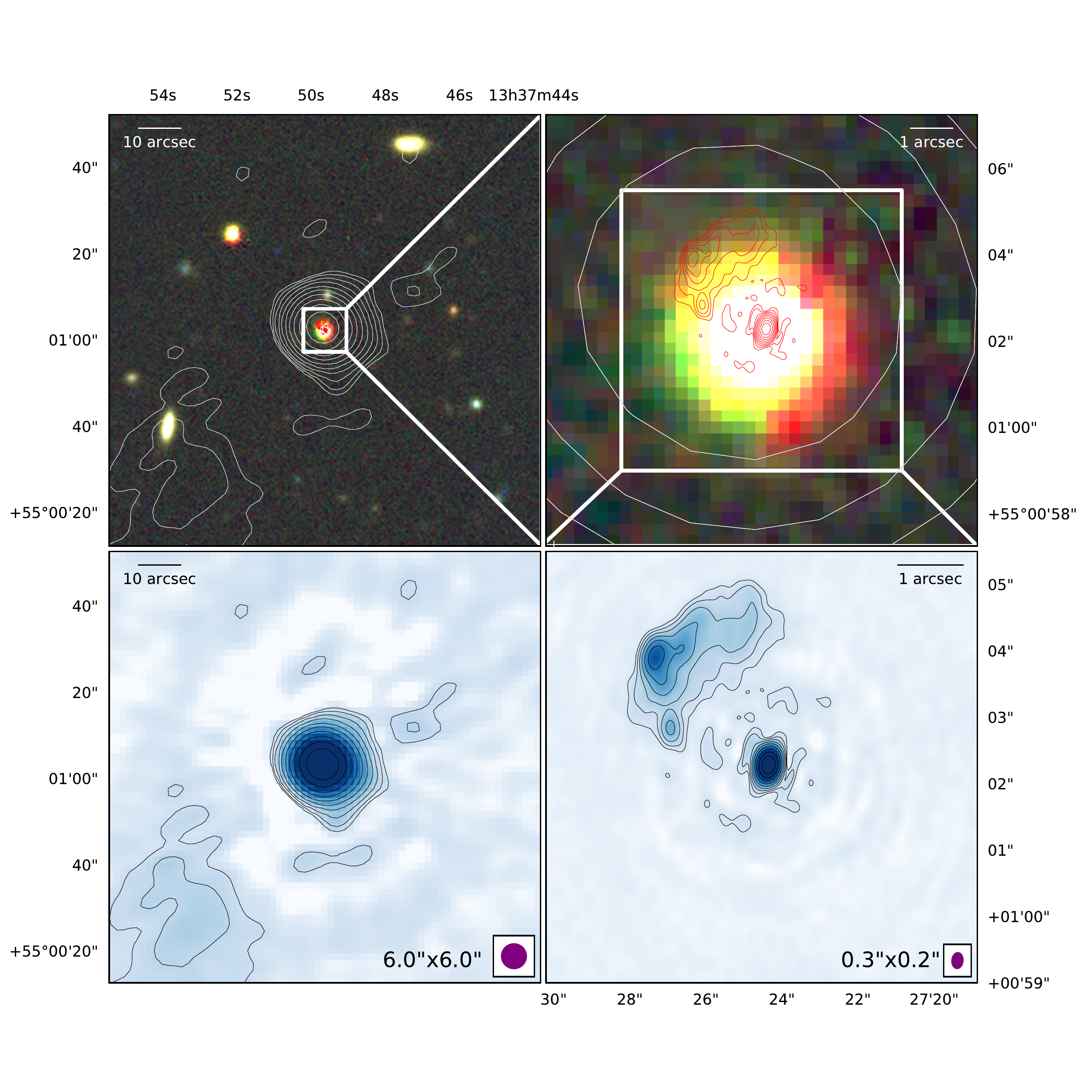}
\caption{\label{fig:cal} Delay calibrator source, ILTJ1337+5501. The background rgb image in the top panels is bands $g,r,z$ from the Legacy Surveys (\href{http://legacysurvey.org/viewer/}{http://legacysurvey.org/viewer/}). The white contours in the two top panels are from the DR2 processing of this field, while the red contours are the LOFAR-VLBI imaging of this source. The bottom left panel shows the DR2 image and contours, while the bottom right panel shows the LOFAR-VLBI image and contours. All contours start at 10$\sigma$ and increase by $\sqrt{2}$ until either the data maximum is reached or nine levels have been drawn. The restoring beam for the radio images are shown in the bottom two plots.  }
\end{center}
\end{figure}

It is typical practice in VLBI to correct the amplitudes on longer timescales once the phase correction is largely complete. After applying the final dTEC/\difmap\ self-calibration solutions, we start the next part of the self-calibration by solving for complex gains using a solution interval of one hour. This improves the image fidelity almost immediately, as can be seen from the first panel in the bottom row of Figure~\ref{fig:dcselfcal}. After two more rounds, there are still spoke-like artefacts in the image that imply that a further short-timescale correction is needed. We end with a final round of complex gain self-calibration with a solution interval of five minutes. In the final image we achieve an rms of 92$\,\mu$Jy$\,$bm$^{-1}$ and a beam size of 0.25\sarc $\times$ 0.18\sarc . The delay calibrator appears to actually be two sources: a bright, compact source associated with a galaxy, and what may be a background wide angle tail galaxy. The radio contours are overlaid on an optical $rgb$ image in Figure~\ref{fig:cal}. 

The image noise properties in Figure~\ref{fig:cal} are not constant across the image. The dynamic range, defined here as the ratio of peak flux density to the radial profile of the rms image noise, around the bright core increases from $\sim$1,000 to $\sim$6,400 moving from $<$2\sarc\ to $>$7\sarc\ away from the location of peak flux density in the image. There are likely several contributing factors here, including artefacts associated with a nearby bright source, despite bandwidth and time smearing from averaging in the direction of the calibrator. Another contribution likely comes from the lack of short baselines (as the core stations have been combined), which means that diffuse emission may not be well-modelled. Both of these issues can contribute to imperfect calibration. The calibration itself may also contribute to the image noise as we have solved for dispersive delays in the form of dTEC but not non-dispersive delays (clock) as these effects are expected to be small. We expect further testing and development of the pipeline on a wide variety of calibrator sources in a large range of ionospheric conditions to help resolve these issues in the future. 

\subsection{LBCS calibrators: dTEC and phase referencing}
\label{subsec:phaseref}
One technique often used in VLBI is phase referencing. This technique is used when the science target is weak and not appropriate for self-calibrating. Following this principle, we preformed self-calibration on a nearby calibrator and transferred to the target before imaging.  The solution transferral is limited by the spatial cross-section of the coherence volume (i.e. the cross-section of the solid angle of the coherence volume that lies in the sky plane), which is both wavelength dependent (larger for longer wavelengths) and time dependent. For LOFAR, this volume is effectively set by the ionospheric conditions (which can decrease the theoretical coherence volume in the absence of atmosphere), assuming the solutions are found with a perfect model of the calibrator source. 

\begin{figure*}
\centering
\includegraphics[width=0.49\textwidth]{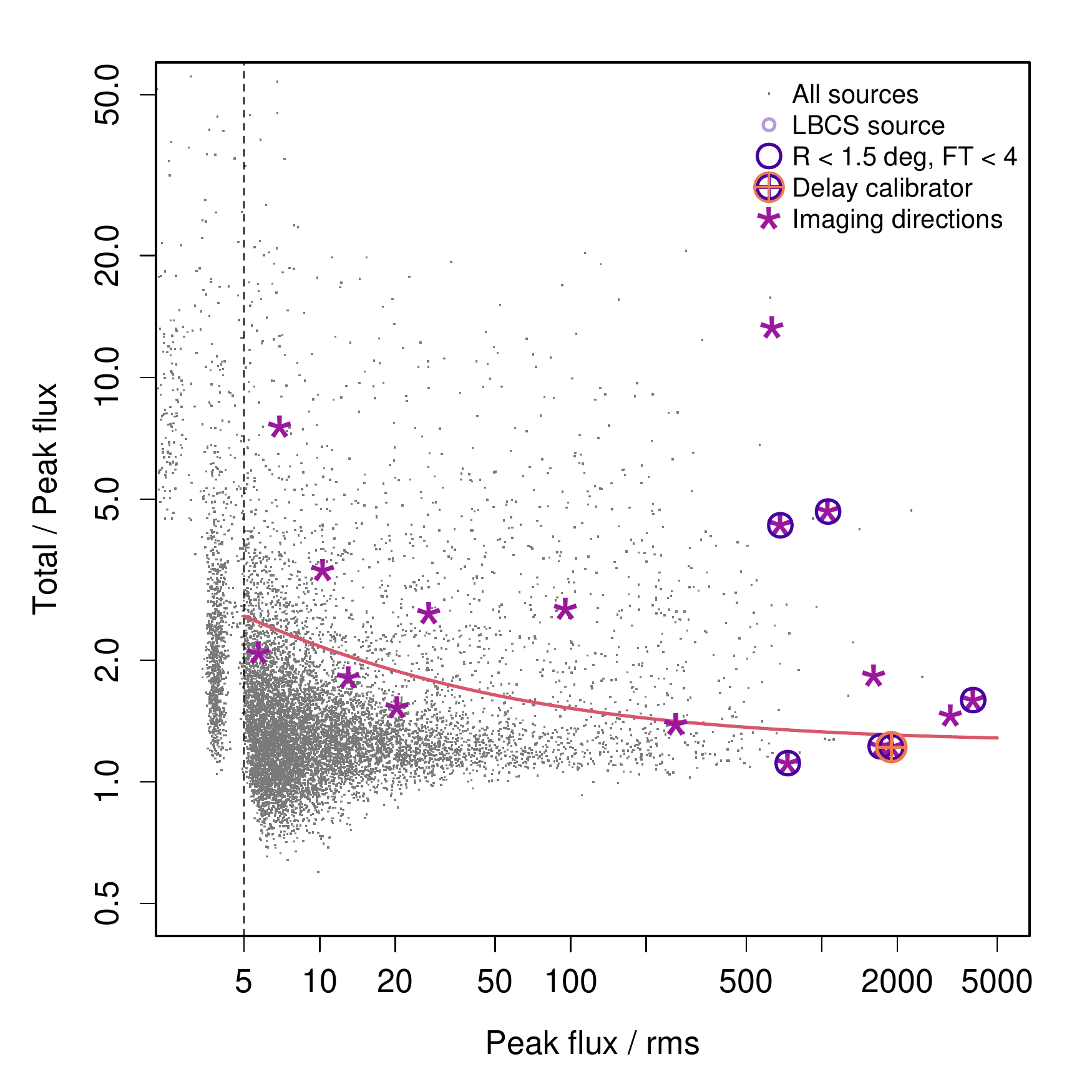} \hfill
\includegraphics[width=0.49\textwidth]{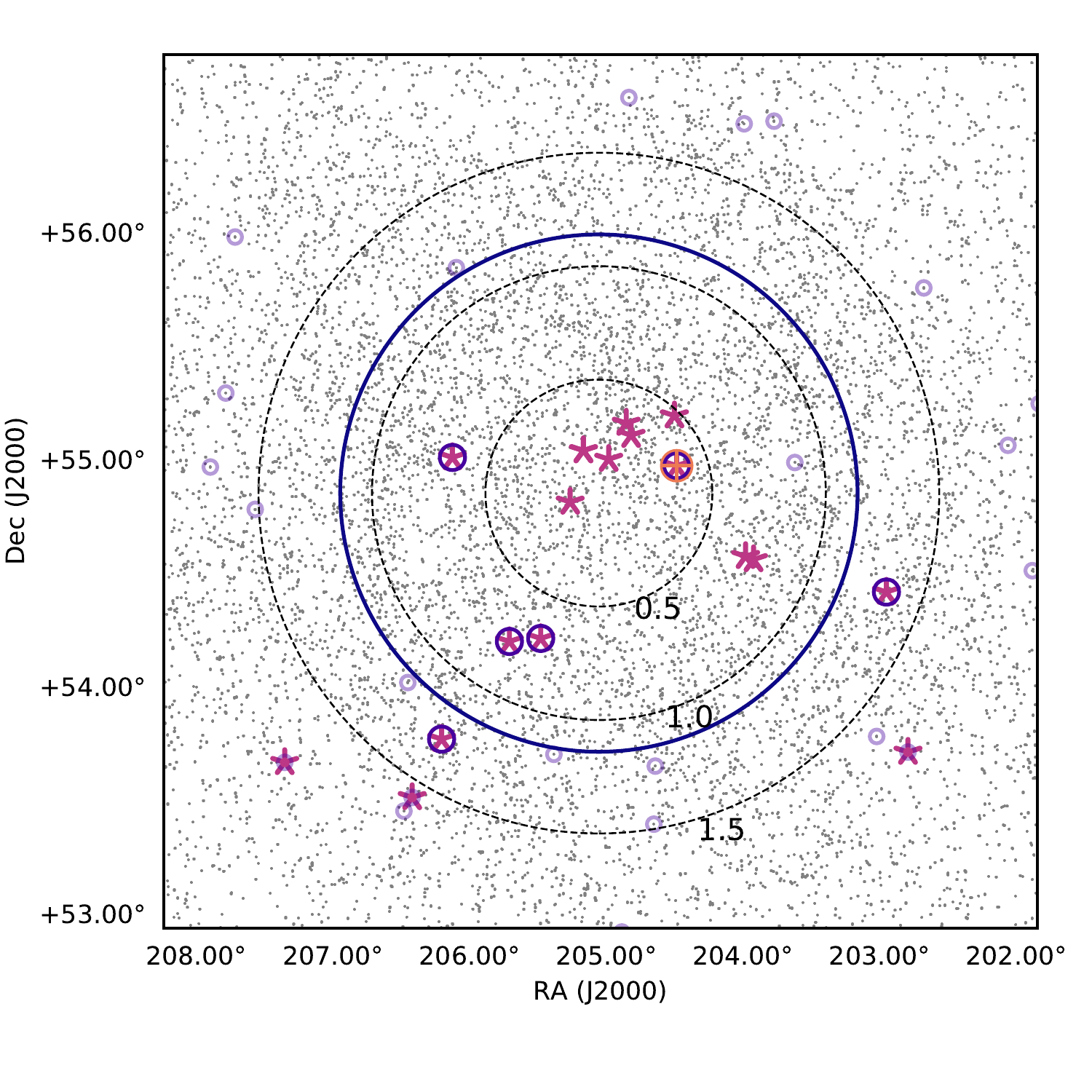}
\caption{\label{fig:catalog} Source properties for the field. The left panel shows the LoTSS total-to-peak flux density ratio as a function of peak flux density/rms noise. The solid orange curve shows the division between resolved and unresolved sources \cite[as in][Section 3.1]{shimwell_lofar_2019}. The vertical dashed black line represents a signal-to-noise ratio of 5. The right panel shows the sky distribution of the sources in the centre of the field, with dotted lines at 0.5, 1.0, and 1.5 degree radii. The solid blue circle at a radius of 1.14 denotes where the intensity losses are expected to be 50 percent due to bandwidth and time smearing. In both panels, all sources detected in the \ddfp\ catalogue are plotted as grey points, with LBCS sources marked with purple circles. The darker, larger purple circles show LBCS sources within 1.5 degrees of the phase centre and having a combined FT Goodness value $>4$. The delay calibrator is marked with an orange cross-hair. Directions of interest (DOIs) are marked by the star symbols.}
\end{figure*}

Phase referencing can also be built up incrementally through multiple calibrators. This is important for high-resolution imaging with LOFAR as the field of view is $\sim$5 square degrees and the delay calibrator may be substantially distant from the desired target. In P205+55 there are five LBCS calibrators in addition to the delay calibrator within 1.5 degrees of the phase centre, and having reasonably high combined FT Goodness statistics ($>4$). These LBCS calibrators are all $\sim$1 degree away from the delay calibrator, which is typically what we would expect from the LBCS sky density. These sources are some of the brightest in the field, ranging from 0.43 Jy to 1.8 Jy. The largest distance between these LBCS calibrators and the delay calibrator is 1.6 degrees; more detailed information is given in Table~\ref{tab:doi} and Figure~\ref{fig:catalog}.

\begin{figure*}
\centering
\includegraphics[width=\textwidth]{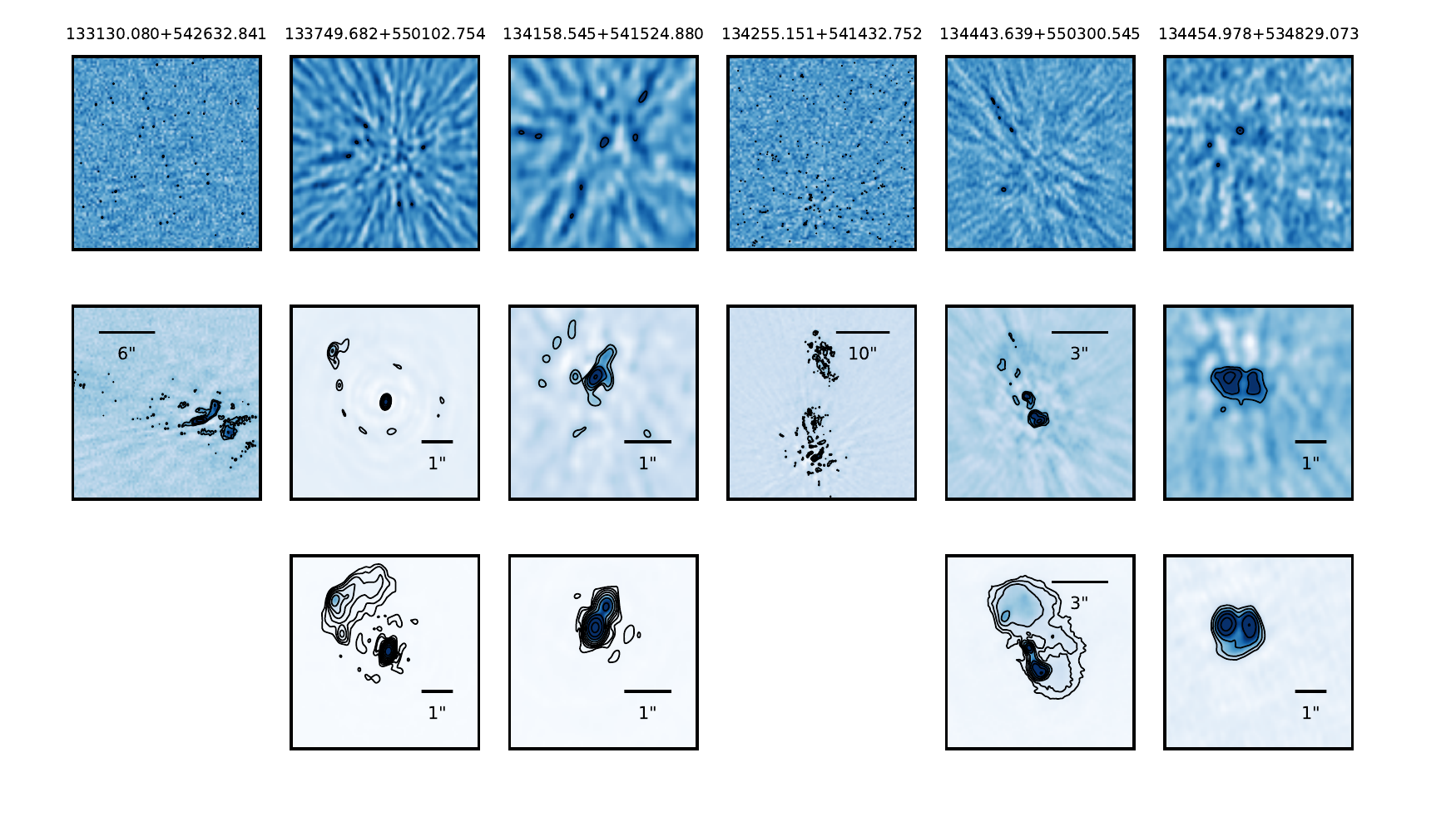}
\caption{\label{fig:lbcscals} Images of the six LBCS calibrators (including the delay calibrator) within 1.5 degrees of the phase centre and having a combined FT Goodness value $>4$. Each column is one LBCS calibrator. An angular scale bar is shown for each column rather than absolute coordinates as the astrometry has not yet been corrected. The image sizes are consistent within a column, but vary from column to column based on the size of the source. The top row shows the DOI for each source with only the dispersive delay from the calibrator applied; we do not expect to see sources here as the data are not phase-calibrated. The middle row shows the DOIs with solutions from the delay calibrator (which is the second column from the left) applied. These solutions include residual dTEC, \difmap\ self-calibration and complex gain solutions. The bottom row shows the DOIs after self-calibration in the same manner as the delay calibrator, starting from models made using the images in the middle row. Two sources with clearly extended emission are not treated as LBCS calibrators and thus do not have a final self-calibrated image in the bottom row of this figure. 
Contours in the middle (bottom) row start at 3$\sigma$ (10$\sigma$) from the image noise and increase by $\sqrt{2}$ to 0.8$\times$(data maximum value), with a limit of the five lowest contours plotted. The colour scale in each image has a linear stretch from $-5\sigma$ to the mean of the contour levels.}
\end{figure*}

Although we have applied the bulk dispersive delay correction to the data, the phases on the international stations are still un-calibrated. We transfer the self-calibration solutions from the delay calibrator to the other five LBCS calibrators and image with \wsclean . Specifically, we transfer the best dTEC, \difmap , and complex gain solutions. The uncorrected and corrected images are shown in the top and middle rows of Figure~\ref{fig:lbcscals}. Two of the sources, ILTJ133130.080+542632.841 and ILTJ134255.151+541432.752, are much larger in size than the other LBCS calibrators (and the position of ILTJ133130.080+542632.841 appears to have been incorrectly catalogued). We set these two sources aside for the moment and focus only on the compact LBCS calibrators, where the self-calibration will be better constrained. The starting point for these sources are the images of the sources themselves, rather than the delay calibrator, where we started with a point source.

Following a similar procedure as for the delay calibrator, we perform two rounds of residual dTEC+\difmap\ self-calibration, followed by three rounds of slow (1 hour) complex gain self-calibration and a final fast (5 minutes) complex gain self-calibration. We found that for ILTJ134454.978+534829.073, the final fast calibration reduced the image fidelity drastically, and we take the image from the previous iteration as the final one. This source is 1.3 degrees from the phase centre, and the furthest from the delay calibrator at a distance of 1.6 degrees, and thus it is not surprising that the solutions on timescales of 5 minutes are not applicable across such a large distance. The starting and final images are shown in the middle and bottom rows of Figure~\ref{fig:lbcscals}, respectively. The noise reached in the final images varies per source, but is $\sim100\,\mu$Jy$\,$bm$^{-1}$ except for ILTJ134454.978+534829.073, which has $230\,\mu$Jy$\,$bm$^{-1}$ image noise.

Two sources are within 1.5 degrees of the phase centre, but outside the radius of 1.14 degrees where the intensity losses are $\gtrsim$50 percent due to bandwidth and time smearing. One of these, ILTJ134454.978+534829.073, we are able to self-calibrate to some extent but reach an rms $\sim$2$\times$ that of the bright compact LBCS calibrators closer to the phase centre. The other source, ILTJ133130.080+542632.841 could potentially be improved by phase referencing to an LBCS calibrator source nearer than the delay calibrator, but unfortunately no such calibrator exists, so we do not attempt anything further with it. The only other LBCS source we have not yet calibrated is ILTJ134255.151+541432.752. We will return to this in Section~\ref{sec:results}. 

The LBCS calibrator solutions then form the basis for imaging other DOIs in the field. The residual dTEC solutions are all $< 0.1\,$TECU for the successful LBCS calibrators, even the one 1.6 degrees away from the primary delay calibrator. This implies that the dispersive delays vary slowly across the field of view, and can be applied up to $\sim1.5$ degrees away, while the phases have more limited validity. Phase transferral works for the sources $\sim$1 degree away from the delay calibrator, but may have some residual problems transferring to the more distant calibrator. As the LBCS calibrator solutions are incremental after the delay calibrator solutions are applied, both the solutions from the delay calibrator and the nearest LBCS calibrator to the DOI should be transferred. Ideally the solutions from a nearby calibrator will be enough to image the science target without resorting to self-calibration.  

\subsection{Astrometry and flux density scale}
\label{subsec:astrom}
Typically finding an astrometric solution and fixing the flux density scale for VLBI sources requires information from ancillary data or catalogues. For example, astrometry can be checked and corrected by identifying source(s) in the field with which to match against the known sky positions of sources. The flux density scale can be compared against known tabulated flux densities, or SED extrapolations from data at other frequencies. Here we make use of LoTSS, which has already had its astrometry and flux density scale fixed \citep{shimwell_lofar_2019,williams_lofar_2019}. In this unique situation, we can use the data itself to correct both the astrometry and flux density scale. At the end of the high-resolution self-calibration, we copy the solutions over to a measurement set for the DOI but having the core stations instead of the combined super-station ST001. The solutions for ST001 are copied to each core station while the remote station solutions are applied directly. Filtering the international stations out, we can then image the direction with \wsclean\ parameters mimicking the LoTSS imaging. Both the astrometry and the flux density scale can then be directly compared between LoTSS and this 6\sarc\ resolution image. 
The calibration pipeline for LoTSS performs astrometric corrections on a facet basis, using radio sources cross-matched with the Pan-STARRS catalogue \citep{flewelling_pan-starrs1_2016}. The errors on the offsets have a median of 0.2\sarc\ \citep{shimwell_lofar_2019}. The LoTSS-DR2 flux density scale is aligned with the \cite{roger_low-frequency_1973} flux density scale through cross matching with NVSS and assuming a global scaling factor to align this with 6C. This procedure, which is thought to produce an overall accuracy of order 10 percent, is briefly described in \cite{hardcastle_contribution_2020} and will be detailed in Shimwell et al. in preparation. 

\begin{figure}[ht]
\centering
\includegraphics[width=0.5\textwidth]{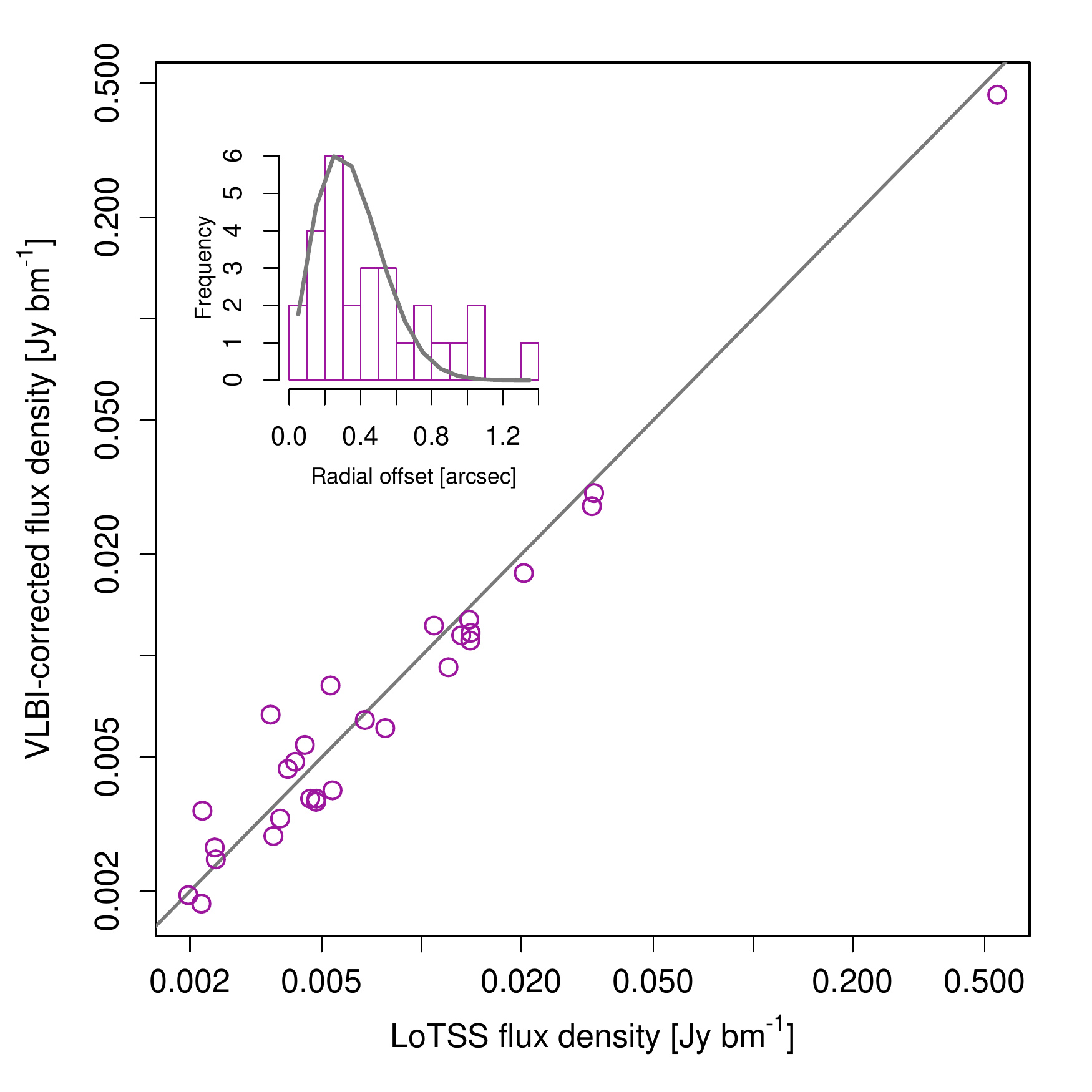}
\caption{\label{fig:astrom} Flux density scale and astrometry checks, using the LoTSS DR2 image as a reference. There are 28 sources cross-matched between the LOFAR-VLBI corrected image and the reference LoTSS DR2 image. The large plot shows the integrated flux densities of these sources plotted against each other, with a line of equal flux density to guide the eye. The inset plot shows the astrometric precision. The histogram is the distribution of radial offsets of the difference in angular positions of the sources, in arcseconds. The grey line is a Rayleigh distribution with parameter $s=0.28$\sarc\ that describes the radial offsets well.  }
\end{figure}

We imaged a region with a 0.22 degree radius (0.43 degrees on a square side) and extracted a cutout from the DR2 mosaic matching this specific area. Running \textsc{PyBDSF} over the two images yields 28 sources that can be cross-matched between the two catalogues with a search radius of 2 arcsec. We use these sources to check both the flux density scale and the astrometry. The total flux densities from the matched sources are plotted against each other in Figure~\ref{fig:astrom}, with a line of equal flux density to guide the eye. The scatter around this line is $\sim$28 percent, which reduces to 14 percent when considering only the 17 isolated point sources. The inset in Figure~\ref{fig:astrom} shows the distribution of the radial offset around zero of the difference between the sky coordinates. To find this, we first determine $\Delta$RA and $\Delta$Dec separately and subtract the average value of each (which is a systematic offset) to centre them on zero. The average values were $\Delta$RA=0.237\sarc\  and $\Delta$Dec=0.288\sarc\ . Systematic offsets are common in VLBI as there is a degeneracy between phase and sky position; often double sources will be artifically re-centred by the calibration solutions so the brightest component is at the phase centre. The true uncertainty in the astrometric accuracy is defined by the distribution of the radial offset from zero. This is a combination of the distributions of two independent parameters (RA and Dec), which is described by a Rayleigh distribution, which is defined by the parameter $s$. We find the radial offsets are well fit with $s=0.28$\sarc\ (see Figure~\ref{fig:astrom}).

\section{Results}
\label{sec:results}
\begin{sidewaysfigure*}
\centering
\includegraphics[width=\textwidth]{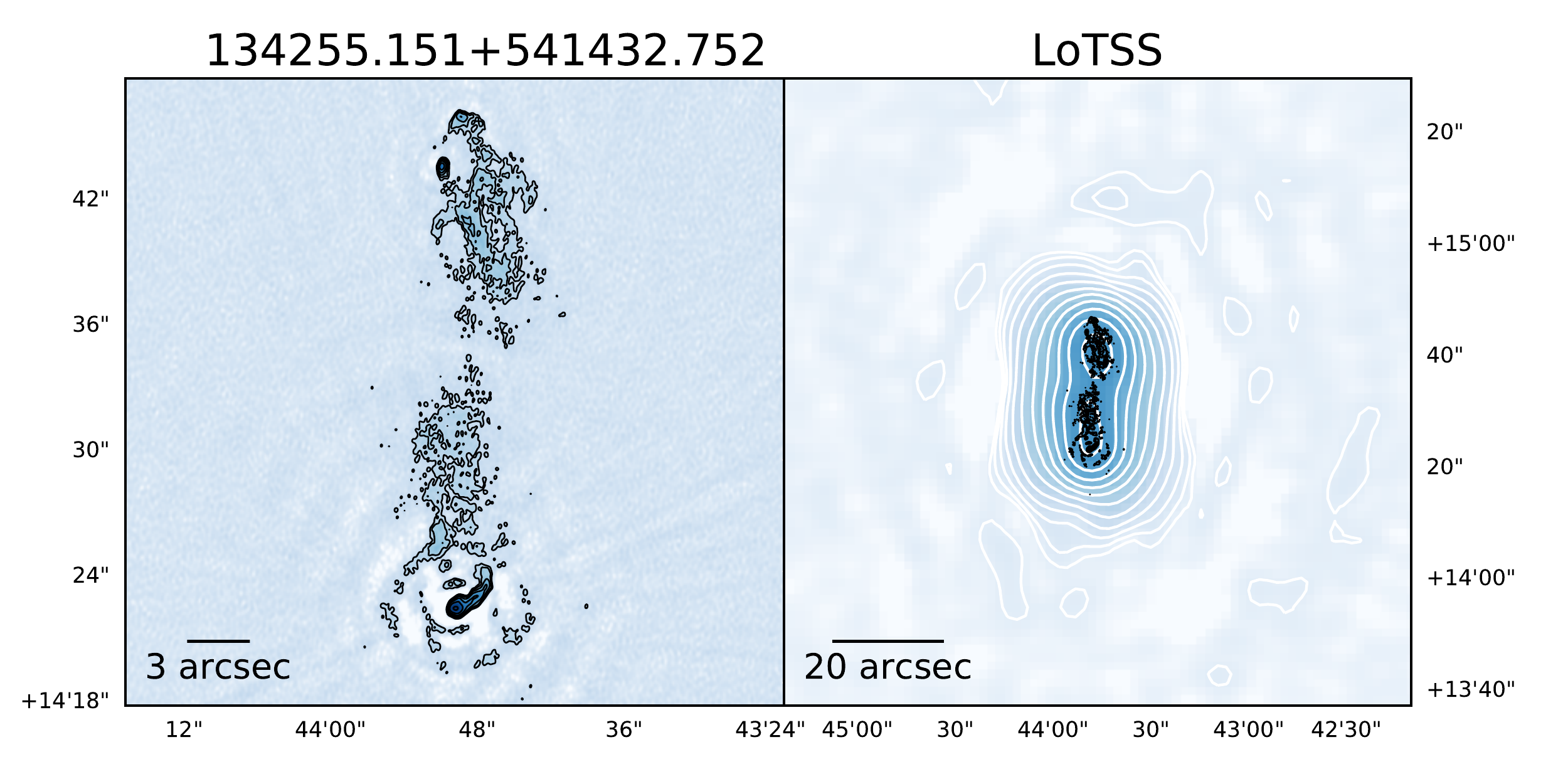}
\caption{\label{fig:frii} Source ILTJ134255.151+541432.752, which was initially selected as an LBCS calibrator but is actually $\sim$20 arcsec in extent. The hotspots are clearly visible, and the core of this galaxy is tentatively detected with 5$\sigma$. The bright compact source in the northern lobe may be an interloping galaxy, as the northern hotspot is visible at the tip of the emission plume. The left panel shows the LOFAR-VLBI image, with a noise level of 99.33$\,\mu$Jy$\,$bm$^{-1}$ and a beam size of 0.29\sarc $\times$ 0.21\sarc\ . The contours start at 5$\sigma$ and increase in steps of $\sqrt{2}$ for a maximum of 5 levels. The colourscale is a log stretch from $-5\sigma$ to the maximum contour level. The right panel shows the 6\sarc\ LoTSS image, with white contours defined in the same way relative to the noise in the 6\sarc\ image. The black contours from the high-resolution image are overlaid. }
\end{sidewaysfigure*}

We have shown how the pipeline works for the primary delay calibrator and other LBCS calibrators in the field, with the exception of  ILTJ134255.151+541432.752. The 6\sarc\ resolution image shows it is a Fanaroff-Riley Type 2 \citep{fanaroff_morphology_1974} source $\sim$20\sarc\ in extent. Fortunately, it is only 8.3 arcmin away from the nearby calibrator ILTJ134158.545+541524.880, offering an alternate route to calibrate and image this source: We simply transfer both the delay calibrator solutions, which were used as the basis for the self-calibration of ILTJ134158.545+541524.880, and then the self-calibration solutions from ILTJ134158.545+541524.880 itself. The resulting image of ILTJ134255.151+541432.752 is presented in Figure~\ref{fig:frii}. We have not performed any self-calibration on the source itself, to showcase the phase referencing technique. The noise achieved is 99$\,\mu$Jy$\,$bm$^{-1}$ with a beam size of 0.29\sarc $\times$ 0.21\sarc\ . 

\begin{figure*}
\centering
\includegraphics[width=\textwidth]{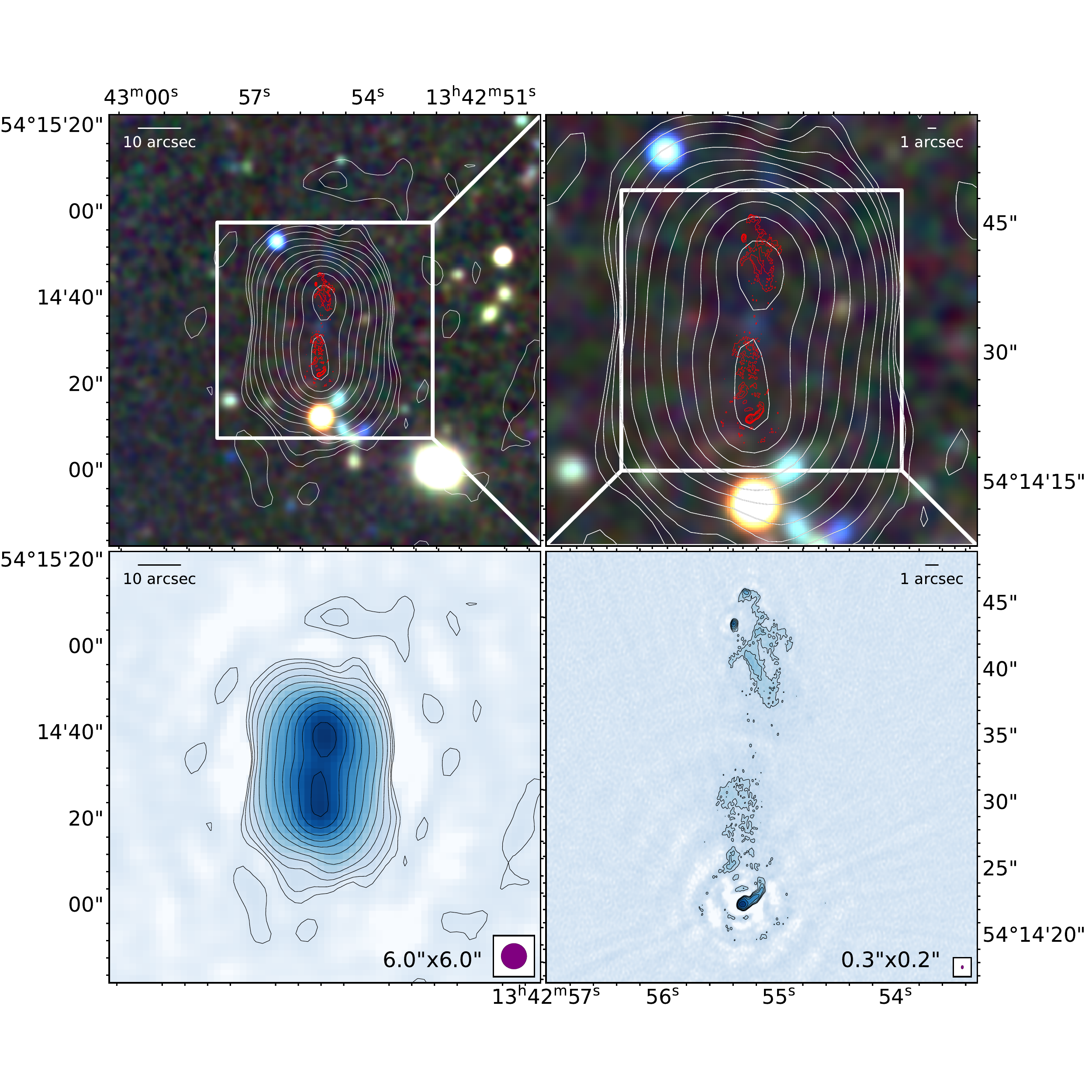}
\caption{\label{fig:134255optical}  Source ILTJ134255.151+541432.752. The background rgb image in the top panels is bands $g,r,z$ from the Legacy Surveys as in Figure~\ref{fig:cal} but here it is smoothed with a 2D Gaussian kernel with width=0.8\sarc\ . The white contours in the two top panels are from the DR2 processing of this field, while the red contours are the LOFAR-VLBI imaging of this source. The bottom left panel shows the DR2 image and contours, while the bottom right panel shows the LOFAR-VLBI image and contours. The LoTSS contours start at 10$\sigma$ and increase by $\sqrt{2}$ until either the data maximum is reached or nine levels have been drawn, while the LOFAR-VLBI contours start at 7$\sigma$ and increase by $\sqrt{2}$ until five levels have been drawn. The restoring beam for the radio images are shown in the bottom two plots.}
\end{figure*}

We also show ILTJ134255.151+541432.752 overlaid on optical data in Figure~\ref{fig:134255optical}. The optical data consist of $g,r,z$ bands from the Legacy Surveys, and is smoothed with a 2D Gaussian kernel (width=0.8\sarc ) to bring out the detection of the faint optical host, which is clearly identified in the WISE 3.5 $\mu$m band. This optically faint galaxy has a photo-$z$ of 1.2 in the Legacy Surveys Data Release 8 database. There are two possibilities for the radio core in this image, which could be confirmed by aligning the source with higher-frequency images. We searched for higher-frequency interferometric radio images, but although this source is detected in FIRST the resolution is not high enough to distinguish a core. 

\begin{figure*}
\centering
\includegraphics[width=0.49\textwidth]{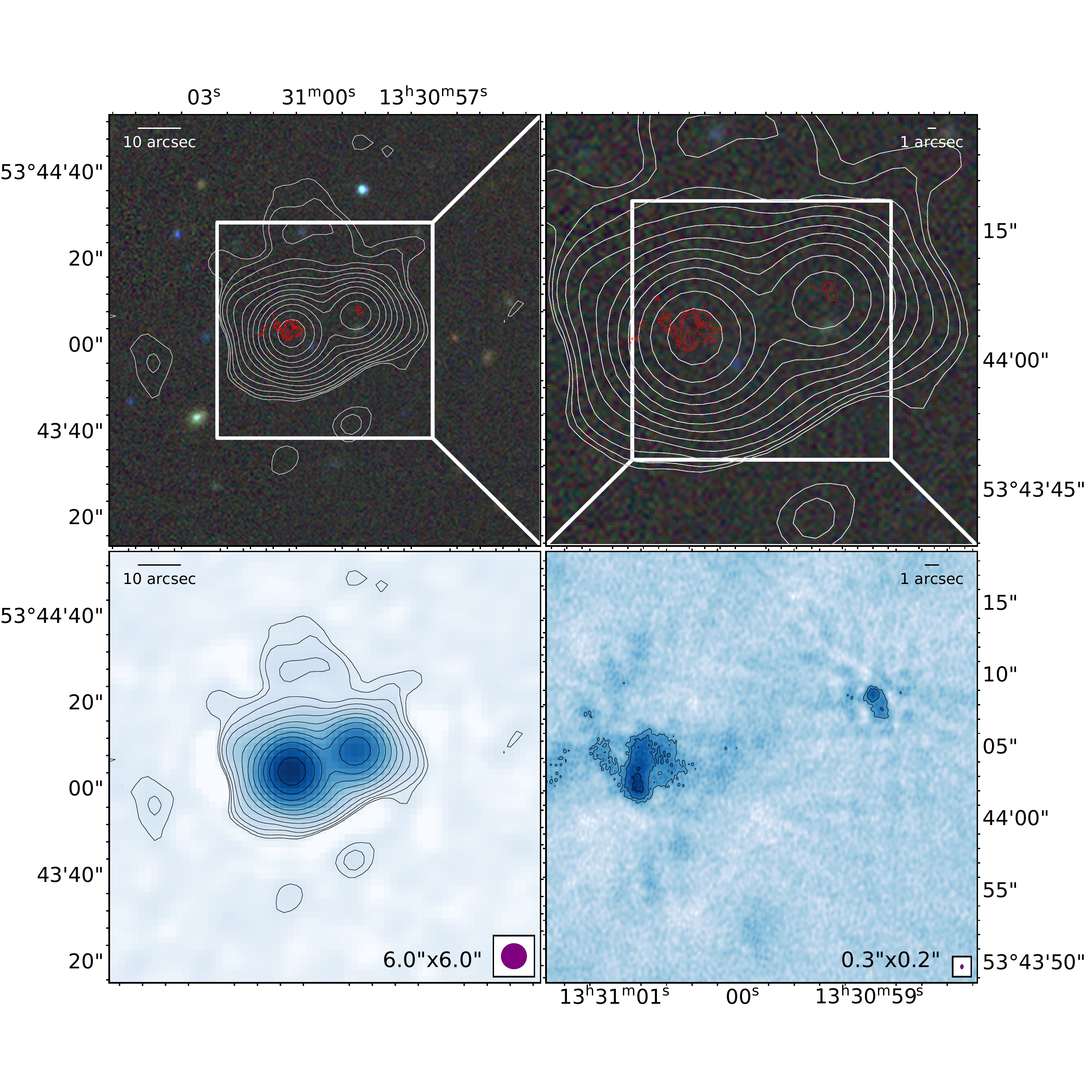}
\includegraphics[width=0.49\textwidth]{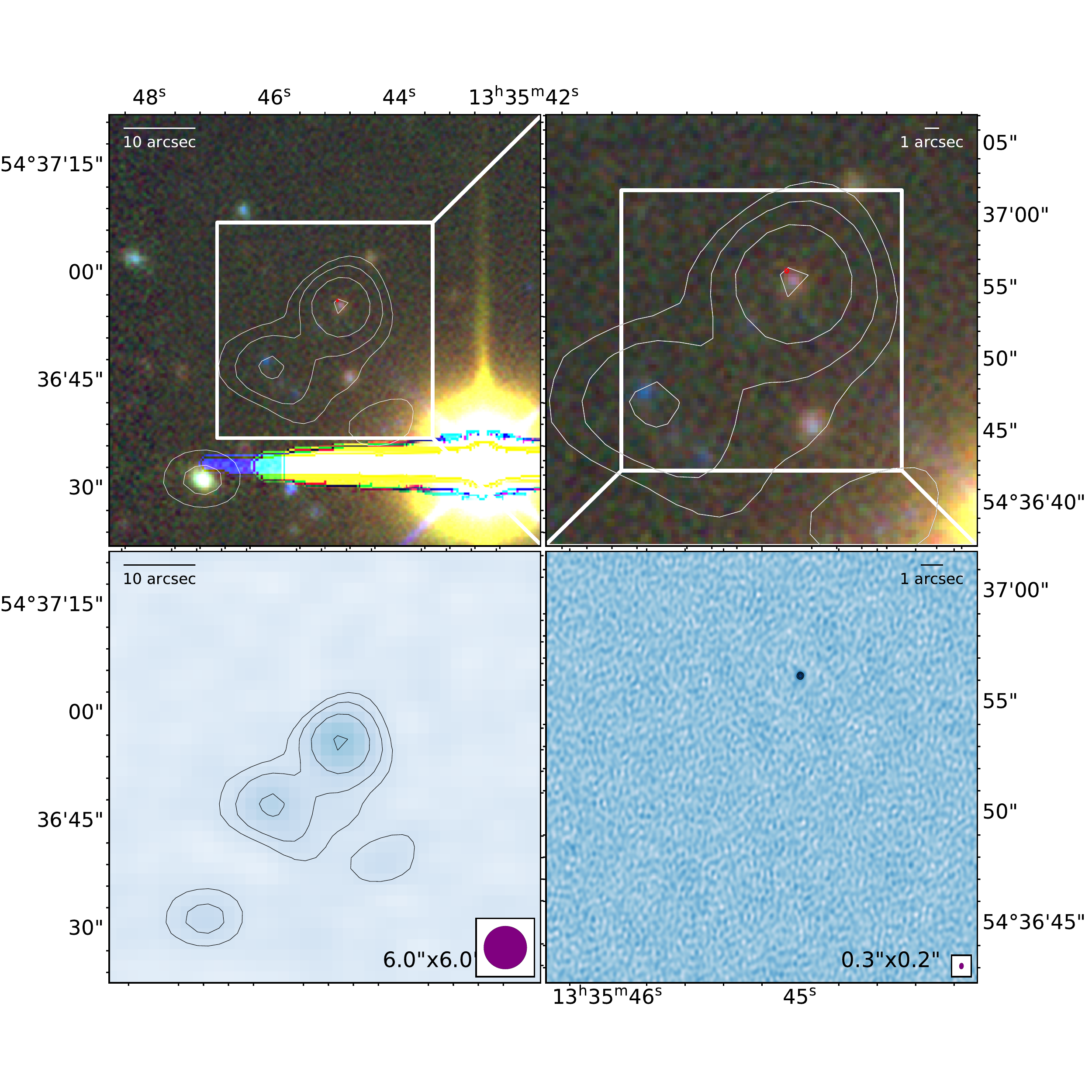}
\includegraphics[width=0.49\textwidth]{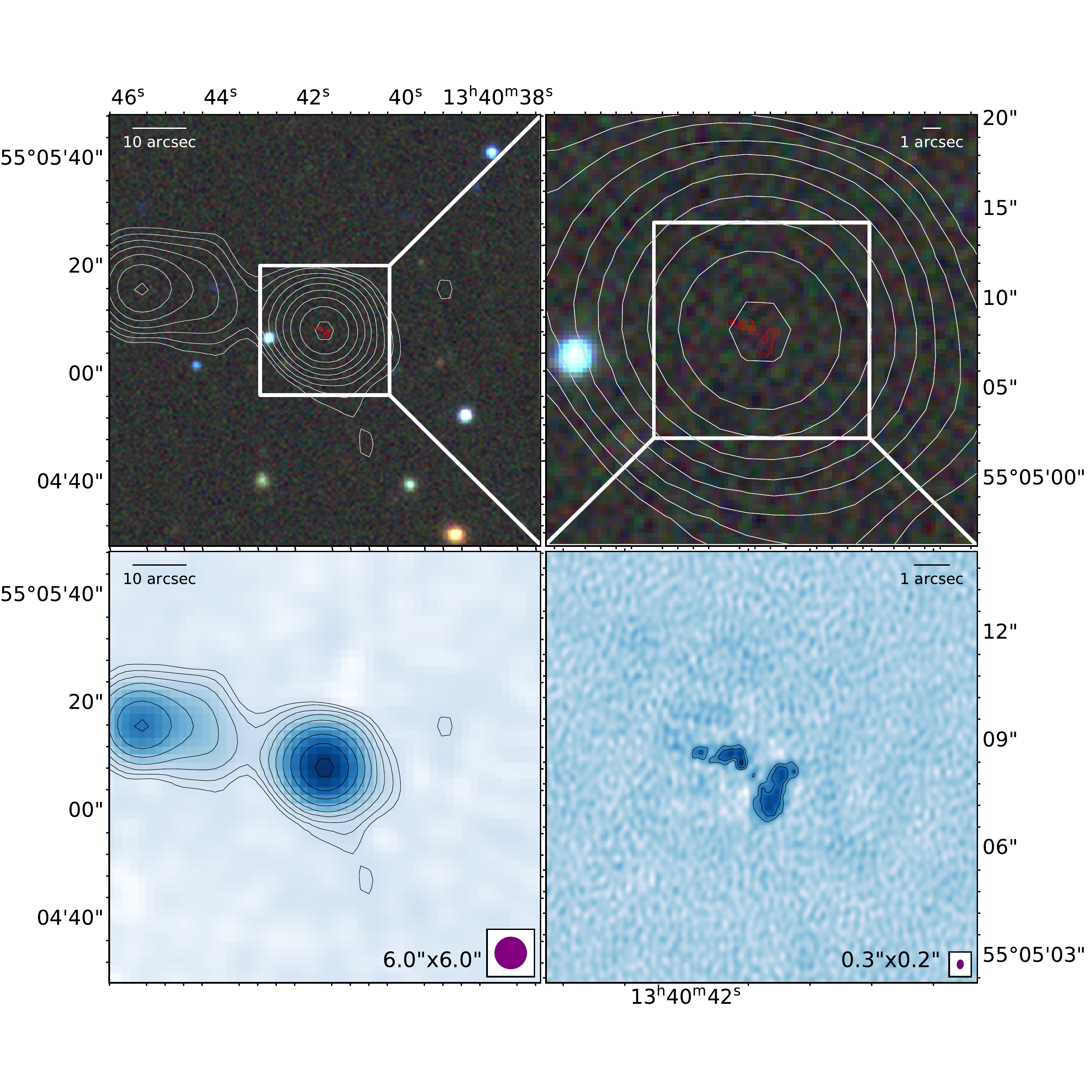}
\includegraphics[width=0.49\textwidth]{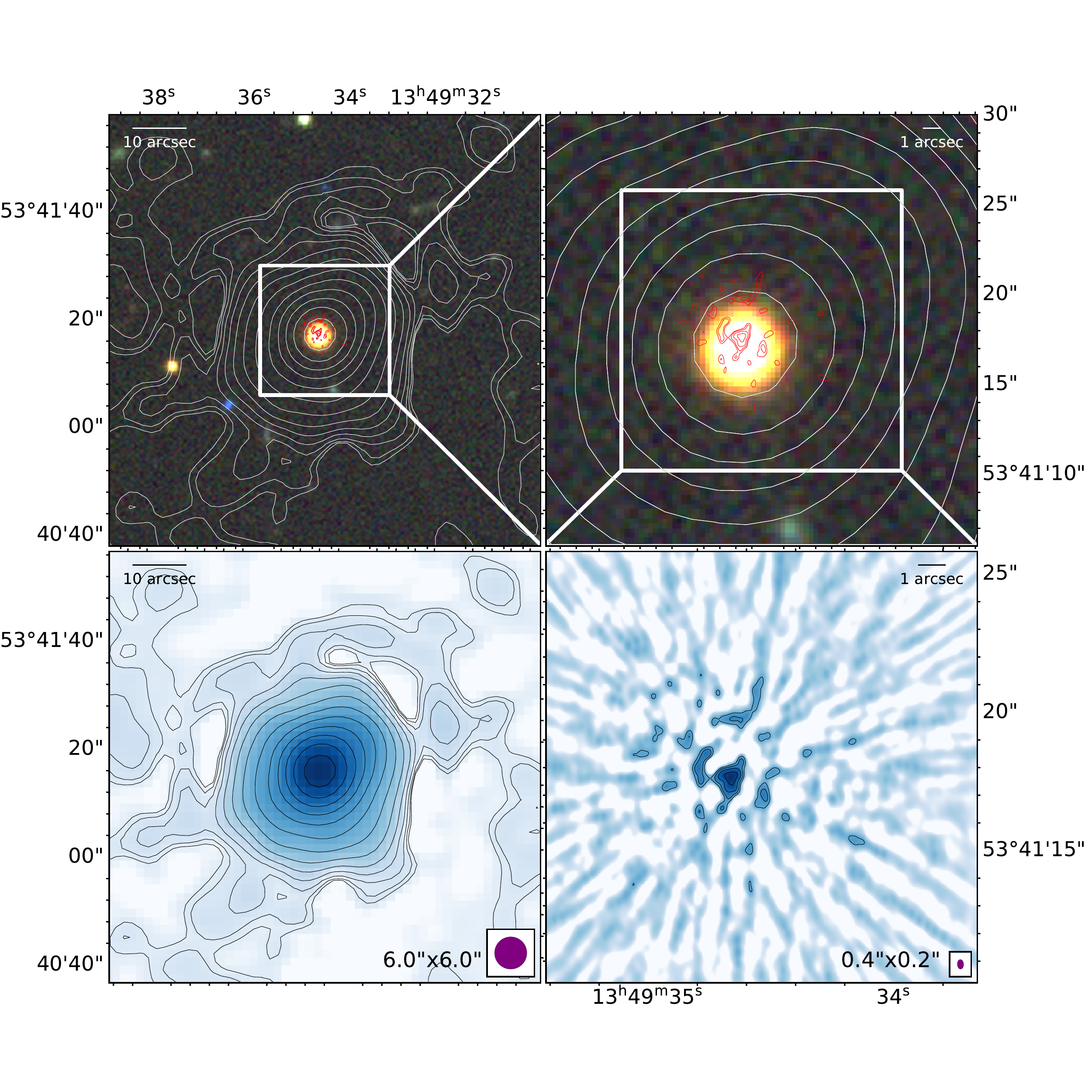}
\caption{\label{fig:dois} Selection of other sources in the field. The background images and contours are defined as in Figures~\ref{fig:cal} and \ref{fig:134255optical}, with the exception of the source in the bottom right panel. This is the brightest source in the field (2.5 Jy) and the high-resolution contours start at 50$\sigma$. The colour scale in all images is based on the noise in the image, and runs with a log stretch from -5$\sigma$ to the maximum contour level.}
\end{figure*}

To further demonstrate the pipeline we image in a number of different directions. These were selected semi-randomly, to represent sources with a wide range of S/N in the LoTSS catalogue, with some resolved and some unresolved (at 6\sarc ) across the range of S/N, and at varying radii from the phase centre. This is shown in Figure~\ref{fig:catalog}. We image these non-LBCS sources by applying the solutions from the delay calibrator and the nearest LBCS calibrator (if it is not the delay calibrator). These are shown in Figure~\ref{fig:dois} and image noises and beam sizes are are given in Table~\ref{tab:doi}. We find the average image noise to be $\sim$90$\,\mu$Jy$\,$bm$^{-1}$ and the average beam size to be 0.3\sarc $\times$0.2\sarc\ . There were some non-detections; these could be due either to source structure if there is no compact emission to be detected or distance from the calibrator(s). Although further self-calibration of some of these brighter sources is possible, we do not perform any self-calibration on these science directions. In the case of faint or very complex science targets, the best practice is to phase reference from a nearby calibrator and then image. This avoids self-calibration converging to an incorrect source model, which may not always be obvious, and can be particularly hard to catch in an automated pipeline. We suggest trying different starting models to see if the self-calibration converges to the same final source structure.

Although in the case of ILTJ134255.151+541432.752 we found that transferring the self-calibration solutions from a nearby calibrator in addition to the delay calibrator improved the image fidelity, the secondary calibrator solutions do not always help for more distant targets. This implies that the solutions of secondary calibrators in this field may not be valid over larger distances. Before investigating this further, we aim to improve the initial delay calibration, which may have downstream consequences, and thus we defer this to a future paper. We therefore caution the user to check their solution transfer carefully. From Table~\ref{tab:doi}, it is clear that the noise increases with the distance from the phase centre. As discussed in Section~\ref{subsec:furtherselfcal}, there are likely residual contributions to the noise on the self-calibrated LBCS calibrator(s), which will be transferred to these sources. On top of this, any DDEs that are different between the calibrator(s) and source imaging directions will contribute to the noise as we do not self-calibrate in the imaging directions. These will be reduced with further development of the pipeline, by implementing, for example, solution screens that can be interpolated between directions. Processing different fields with a range of sources and ionospheric conditions will also help disentangle the residual noise contributions from calibration and from radial field of view limitations. The non-detected sources are all fainter, lower signal-to-noise sources, which are already resolved at 6\sarc\ and therefore may not have much compact emission. 

\begin{table*}
\caption{\label{tab:doi}Directions of interest to demonstrate the automated self-calibration. The radius is distance from phase centre. No beam size is given for sources that are non-detections in the high-resolution images.}
\begin{tabular}{lccccccc}
Source\_id & LBCS & $S_{\textrm{int}}$ & $S_{\textrm{peak}}$ & radius & distance to cal & noise & beam \\ 
 & & [mJy] & [mJy] & [$^{\circ}$] & [$^{\circ}$] & [$\mu$Jy$\,$bm$^{-1}$] & [arcsec] \\[2pt] \hline \\[-6pt] 
ILTJ134934.536+534118.042 &    & 2514.99  & 1727.51 & 1.82 & 2.17  & 315  & 0.34$\times$0.23 \\
ILTJ134443.639+550300.545 & Y & 1761.70 & 1105.76 & 0.66 & 0.99 & 120 	& 0.26$\times$0.19 \\ 
ILTJ134255.151+541432.752 & Y & 1472.67 & 341.66  & 0.76 & 1.07 & 99   & 0.29$\times$0.21 \\ 
ILTJ134545.624+533255.358 &    & 1423.88 & 107.53  & 1.57     &  1.87    & 216  & 0.34$\times$0.23 \\
ILTJ133100.392+534403.681 &   & 1315.70 & 720.10  &  1.78    & 1.62     & 176.41 	& 0.34$\times$0.24 \\
ILTJ133130.080+542632.841 & Y & 936.36 & 200.90   & 1.34 & 1.08 & 138 	& 0.31$\times$0.23 \\ 
ILTJ133749.682+550102.754 & Y & 667.33 & 547.63   & 0.36 & 0.00 & 92  	& 0.25$\times$0.18 \\ 
ILTJ134454.978+534829.073 & Y & 496.94 & 405.78   & 1.29 & 1.59 & 230 	& 0.31$\times$0.24 \\ 
ILTJ134158.545+541524.880 & Y & 429.82 & 386.32   & 0.69 & 0.97 & 100  & 0.27$\times$0.19\\ 
ILTJ134041.754+550508.289 &   & 90.06  & 64.94      & 0.20     &  0.42    & 86   & 0.27$\times$0.19\\
ILTJ133922.481+551223.167 &    & 20.03 & 7.49     & 0.33     &  0.29   & 86   & -- \\
ILTJ133955.122+550251.647 &   & 5.31   & 0.70     & 0.16 & 0.30 & 84 	& -- \\ 
ILTJ133545.199+543652.094 &   & 5.10   & 1.95     & 0.70 & 0.50 & 90 	& 0.28$\times$0.20 \\  
ILTJ133753.050+551416.717 &   & 2.71   & 0.81 	 & 0.48 & 0.22 & 87	& -- \\ 
ILTJ134105.957+545131.449 &   & 2.19   & 1.43 	 & 0.13 & 0.50 & 84 	& 0.27$\times$0.19 \\ 
ILTJ133530.773+543559.236 &   & 1.73   & 0.96		 & 0.74 & 0.53 & 90	& -- \\ 
ILTJ133913.546+550910.591 &   & 0.81   & 0.39		 & 0.29 & 0.24 & 87	& -- \\ 
\end{tabular}
 \end{table*}

\section{Summary and future work}
\label{sec:conclusions}
We have presented a successful calibration strategy to produce images at sub-arcsecond resolution using the entire International LOFAR Telescope, using a typical LoTSS pointing. Based on previous successful work, this calibration strategy handles the unique challenges of LOFAR imaging by adapting traditional VLBI techniques. The strategy has been implemented in the automated pipeline described in depth in this paper. The pipeline is publicly available, although the final self-calibration of the target is currently expected to be adjusted by the user. 

We provided guidance on the observing strategy, which for targeted observations should place the science target at the phase centre. We suggest using the LOFAR Surveys KSP strategy of having the Radio Observatory only average to 1 sec and 16 channels per subband, to preserve the field of view against bandwidth and time smearing. The standard flux density calibrators should be given preference in order of most to least compact, although appropriate high-resolution models exist for the more complex sources. 

The overall calibration strategy and details on the pipeline processing were presented. In-depth documentation on practical configuration of the pipeline is also provided at \url{https://lofar-vlbi.readthedocs.io/en/latest/}. We discussed the post-pipeline processing steps, including solution referencing amongst sources. We find that the dTEC varies by less than 0.1 TECU across the field after removing the bulk dispersive delay, and that phases can successfully be transferred over $\sim$1.5 degrees. Based on the source density of LBCS and the spacing of LoTSS pointings (separation of 2.3 degrees), this means the entire area of LoTSS should be accessible. As these steps were based on the particular sources and their locations relative to each other in the field, they have not yet been included in the pipeline. Although we have tested it on HBA data, the strategy is similar enough to that used by \cite{morabito_lofar_2016} that it should in principle work for LBA data as well.

The final images reach $\sim90\,\mu$Jy$\,$bm$^{-1}$ noise with an average restoring beam of 0.3\sarc $\times$0.2\sarc\ . We presented the images of the delay calibrator, secondary LBCS calibrators, and a few sources in the field. For the delay calibrator, we demonstrated how the astrometry and flux density scale can be checked and, if necessary, corrected, using LoTSS as a reference. We found good agreement in the flux density scale, with only $\sim$5 percent scatter around a line of equal total flux density. The astrometric accuracy for this source had an overall shift of $<0.3$\sarc\ with individual sources having scatter defined by a Rayleigh distribution with $s=$0.28\sarc\ . 

Although this pipeline is a massive step forward in making LOFAR high-resolution more accessible to non-expert users, there is still more to develop. Several tools have become available recently (e.g. dispersive-delay fringe-fitting) that could help improve the pipeline, and techniques have been developed that should improve the initial delay calibration. We also need to test the pipeline on a variety of fields as the results presented here are dependent on the ionospheric conditions for this particular observation, the source properties of the specific LBCS calibrators in this field, and the sky distribution of sources. Once we are certain that the pipeline is robust in most situations, it can then be optimised for efficiency as it currently takes $\sim$7,000 core hours + $N\times$(250 core hours) to process $N$ sources. This will include work to ensure that the pipeline can be integrated seamlessly with the overall efforts for LoTSS data processing, which include wide-field LOFAR-VLBI techniques. 

\begin{figure}
\centering
\includegraphics[width=0.5\textwidth]{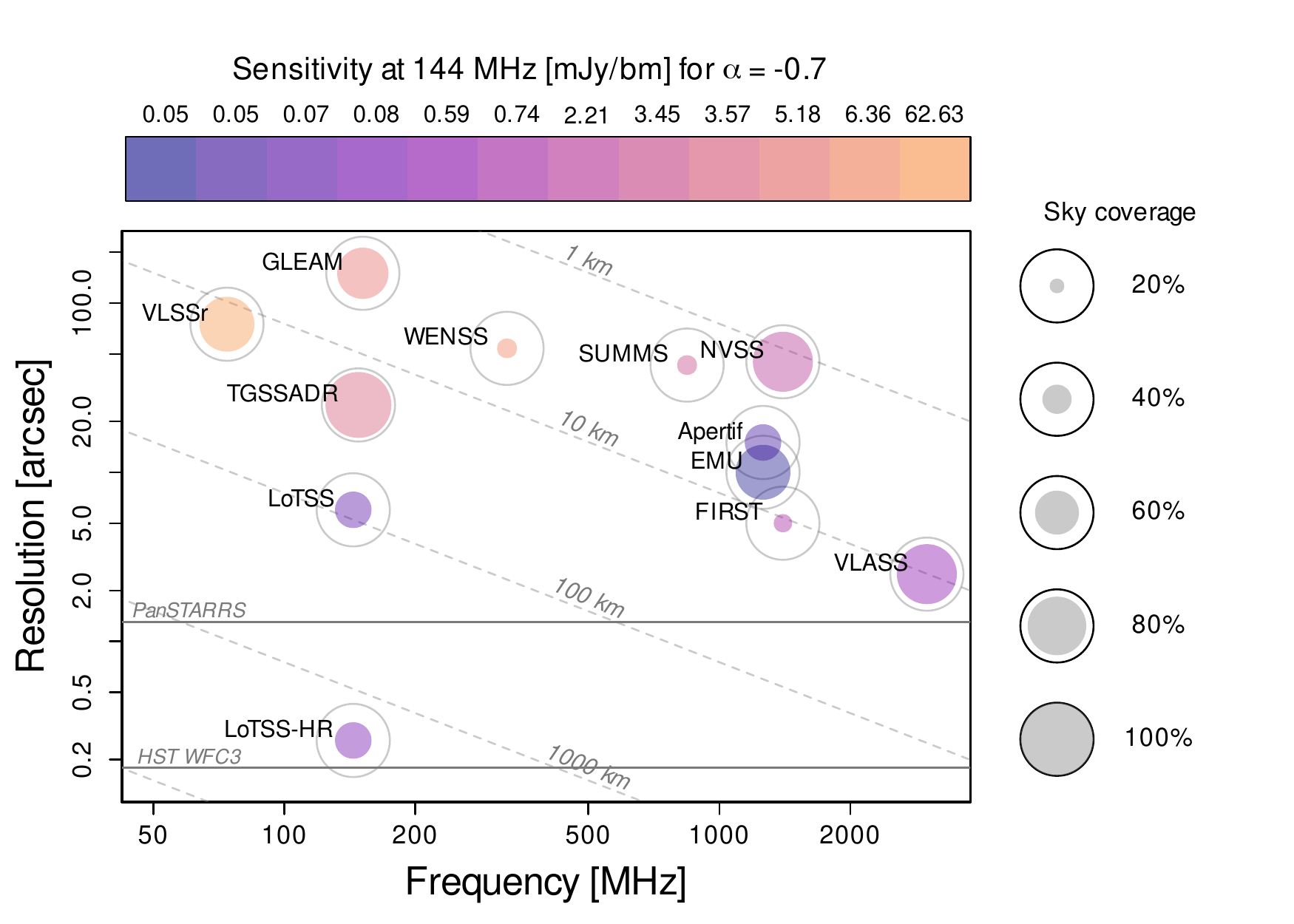}
\caption{\label{fig:surveys} Comparison of resolution, sensitivity, and sky coverage of different radio surveys. The logarithmic x- and y-axes are frequency and resolution, respectively. The colour scale shows the relative survey sensitivity, having scaled the sensitivity of each survey to $\nu=144\,$MHz assuming a typical synchrotron spectral index of $\alpha=-0.7$. The circle outlines show 100\% sky coverage, while the size of the coloured circles correspond to the percentage of the sky that will be covered by the survey; see the key on the right to guide the eye. Horizontal grey lines show typical resolutions of space (Hubble Space Telescope; HST) and ground-based (PanSTARRS) optical telescopes. }
\end{figure}

In the near future, we plan to start post-processing $\sim$100 pointings from LoTSS, which will help us refine and improve the pipeline, in addition to starting to build up the first sub-arcsecond radio survey of the entire northern sky. This survey will be a high-resolution (HR) extension to LoTSS, namely LoTSS-HR. Figure~\ref{fig:surveys} shows a comparison of radio surveys with respect to resolution, sensitivity (projected at 144 MHz, assuming a typical synchrotron spectral index of $\alpha=-0.7$), and sky coverage. LoTSS-HR will be the only survey with resolutions comparable to wide-area optical surveys, and is expected to cover $\sim50$ percent of the entire sky ($>0^{\circ}$ declination). 

 To get an idea of how many sources we will catalogue, we extrapolate from the results here by noticing that all but one source with a peak flux density in LoTSS above 1 mJy have a secure detection. In P205+55, 26 percent of the sources satisfy this flux density limit, and 77 percent of those are unresolved. For P205+55, 3,700 sources are within the 1.14 degree radius where intensity losses are less than 50 percent. We therefore expect we could image around 900 sources in the field. Extrapolating to the entirety of LoTSS, which has 3,300 pointings, this would yield just under 3 million sources. This is expected to be a lower limit as we will continue to optimise the calibration strategy for efficient surveying. The potential for providing morphological information for such a large number of sources, on scales comparable to ancillary optical/NIR data, will unlock the door to a new regime of radio survey science.

\section*{Acknowledgements}
The authors would like to thank everyone who has helped develop the software and pipelines used in this paper. This work would not have been possible without tireless efforts to help update and maintain \prefac\ and \ndppp . The authors are grateful for many useful conversations with K.J. Duncan. This work made use of several different computing resources and we thank the administrators and technicians who have helped us with these resources. This work made use of the University of Hertfordshire high-performance computing facility (\url{http://uhhpc.herts.ac.uk}) and the LOFAR-UK computing facility located at the University of Hertfordshire and supported by STFC [ST/P000096/1]. We would like to thank the Christ Church Research Centre for a grant which provided the disk space necessary to process this data. This work made use of the Dutch national e-infrastructure with the support of the SURF Cooperative using grant no. EINF-262 LKM is grateful for support from the UKRI Future Leaders Fellowship (grant MR/T042842/1). SM acknowledges support from the Government of Ireland Postgraduate Scholarship Programme. EB acknowledges support from the ERC-ERG grant DRANOEL, n.714245. AD acknowledges support by the BMBF Verbundforschung under the grant 052020. JHC acknowledges support from the UK Science and Technology Facilities Council (ST/R000794/1). PNB is grateful for support from the UK STFC via grant ST/R000972/1. JRC thanks the Nederlandse Organisatie voor Wetenschappelijk Onderzoek (NWO) for support via the Talent Programme Veni grant. MJH acknowledges support from the UK Science and Technology Facilities Council (ST/R000905/1). JPM acknowledges support from the Netherlands Organization for Scientific Research (NWO, project number 629.001.023) and the Chinese Academy of Sciences (CAS, project number 114A11KYSB20170054). JM acknowledges financial support from the State Agency for Research of the Spanish MCIU through the ``Center of Excellence Severo Ochoa'' award to the Instituto de Astrof\'isica de Andaluc\'ia (SEV-2017-0709) and from the grant RTI2018-096228-B-C31 (MICIU/FEDER, EU). RJvW acknowledges support from the ERC Starting Grant ClusterWeb 804208. DJS acknowledges support by the German Federal Ministry for Science and Research BMBF-Verbundforschungsprojekt D-LOFAR 2.0 (grant numbers 05A20PB1). LOFAR (van Haarlem et al. 2013) is the Low Frequency Array designed and constructed by ASTRON. It has observing, data processing, and data storage facilities in several countries, that are owned by various parties (each with their own funding sources), and that are collectively operated by the ILT foundation under a joint scientific policy. The ILT resources have benefitted from the following recent major funding sources: CNRS-INSU, Observatoire de Paris and Université d’Orléans, France; BMBF, MIWF-NRW, MPG, Germany; Science Foundation Ireland (SFI), Department of Business, Enterprise and Innovation (DBEI), Ireland; NWO, The Netherlands; The Science and Technology Facilities Council, UK; Ministry of Science and Higher Education, Poland. 

\bibliographystyle{aa}
\bibliography{references}

\begin{thebibliography}{45}
\expandafter\ifx\csname natexlab\endcsname\relax\def\natexlab#1{#1}\fi

\bibitem[{Bonnassieux {et~al.}(2020)Bonnassieux, Edge, Morabito, \&
  Bonafede}]{bonnassieux_decoherence_2020}
Bonnassieux, E., Edge, A., Morabito, L., \& Bonafede, A. 2020, Astronomy and
  Astrophysics, 637, A51

\bibitem[{Bridle \& Schwab(1999)}]{bridle_bandwidth_1999}
Bridle, A.~H. \& Schwab, F.~R. 1999, in Synthesis {Imaging} in {Radio}
  {Astronomy} {II}, Vol. 180, 371

\bibitem[{de~Gasperin {et~al.}(2019)de~Gasperin, Dijkema, Drabent, Mevius,
  Rafferty, van Weeren, Brüggen, Callingham, Emig, Heald, Intema, Morabito,
  Offringa, Oonk, Orrù, Röttgering, Sabater, Shimwell, Shulevski, \&
  Williams}]{de_gasperin_systematic_2019}
de~Gasperin, F., Dijkema, T.~J., Drabent, A., {et~al.} 2019, Astronomy and
  Astrophysics, 622, A5

\bibitem[{de~Gasperin {et~al.}(2018)de~Gasperin, Mevius, Rafferty, Intema, \&
  Fallows}]{de_gasperin_effect_2018}
de~Gasperin, F., Mevius, M., Rafferty, D.~A., Intema, H.~T., \& Fallows, R.~A.
  2018, Astronomy and Astrophysics, 615, A179

\bibitem[{{Fanaroff} \& {Riley}(1974)}]{fanaroff_morphology_1974}
{Fanaroff}, B.~L. \& {Riley}, J.~M. 1974, \mnras, 167, 31P

\bibitem[{Flewelling {et~al.}(2016)Flewelling, Magnier, Chambers, Heasley,
  Holmberg, Huber, Sweeney, Waters, Chen, Farrow, Hasinger, Henderson, Long,
  Metcalfe, Nieto-Santisteban, Norberg, Saglia, Szalay, Rest, Thakar, Tonry,
  Valenti, Werner, White, Denneau, Draper, Hodapp, Jedicke, Kaiser, Kudritzki,
  Price, Wainscoat, Chastel, McClean, Postman, \&
  Shiao}]{flewelling_pan-starrs1_2016}
Flewelling, H.~A., Magnier, E.~A., Chambers, K.~C., {et~al.} 2016, ArXiv
  e-prints, 1612, arXiv:1612.05243

\bibitem[{Greisen(2003)}]{greisen_aips_2003}
Greisen, E.~W. 2003, in {AIPS}, the {VLA}, and the {VLBA}, Vol. 285, 109

\bibitem[{Hamaker \& Bregman(1996)}]{hamaker_understanding_1996}
Hamaker, J.~P. \& Bregman, J.~D. 1996, Astronomy and Astrophysics Supplement
  Series, 117, 161

\bibitem[{{Hardcastle} {et~al.}(2020){Hardcastle}, {Shimwell}, {Tasse}, {Best},
  {Drabent}, {Jarvis}, {Prandoni}, {Rottgering}, {Sabater}, \&
  {Schwarz}}]{hardcastle_contribution_2020}
{Hardcastle}, M.~J., {Shimwell}, T.~W., {Tasse}, C., {et~al.} 2020, arXiv
  e-prints, arXiv:2011.08294

\bibitem[{Harris {et~al.}(2019)Harris, Moldón, Oonk, Massaro, Paggi, Deller,
  Godfrey, Morganti, \& Jorstad}]{harris_lofar_2019}
Harris, D.~E., Moldón, J., Oonk, J. R.~R., {et~al.} 2019, The Astrophysical
  Journal, 873, 21

\bibitem[{Intema {et~al.}(2017)Intema, Jagannathan, Mooley, \&
  Frail}]{intema_gmrt_2017}
Intema, H.~T., Jagannathan, P., Mooley, K.~P., \& Frail, D.~A. 2017, Astronomy
  and Astrophysics, 598, A78

\bibitem[{Intema {et~al.}(2009)Intema, van~der Tol, Cotton, Cohen, van Bemmel,
  \& Röttgering}]{intema_ionospheric_2009}
Intema, H.~T., van~der Tol, S., Cotton, W.~D., {et~al.} 2009, Astronomy and
  Astrophysics, 501, 1185

\bibitem[{Jackson {et~al.}(2016)Jackson, Tagore, Deller, Moldón, Varenius,
  Morabito, Wucknitz, Carozzi, Conway, Drabent, Kapinska, Orrù, Brentjens,
  Blaauw, Kuper, Sluman, Schaap, Vermaas, Iacobelli, Cerrigone, Shulevski, ter
  Veen, Fallows, Pizzo, Sipior, Anderson, Avruch, Bell, van Bemmel, Bentum,
  Best, Bonafede, Breitling, Broderick, Brouw, Brüggen, Ciardi, Corstanje,
  de~Gasperin, de~Geus, Eislöffel, Engels, Falcke, Garrett, Grießmeier,
  Gunst, van Haarlem, Heald, Hoeft, Hörandel, Horneffer, Intema, Juette,
  Kuniyoshi, van Leeuwen, Loose, Maat, McFadden, McKay-Bukowski, McKean,
  Mulcahy, Munk, Pandey-Pommier, Polatidis, Reich, Röttgering, Rowlinson,
  Scaife, Schwarz, Steinmetz, Swinbank, Thoudam, Toribio, Vermeulen, Vocks, van
  Weeren, Wise, Yatawatta, \& Zarka}]{jackson_lbcs_2016}
Jackson, N., Tagore, A., Deller, A., {et~al.} 2016, Astronomy and Astrophysics,
  595, A86

\bibitem[{Jackson {et~al.}(2021)Jackson, Tagore, Deller, Moldón, Varenius,
  Morabito, Wucknitz, Carozzi, Conway, Drabent, Kapinska, Orrù, Brentjens,
  Blaauw, Kuper, Sluman, Schaap, Vermaas, Iacobelli, Cerrigone, Shulevski, ter
  Veen, Fallows, Pizzo, Sipior, Anderson, Avruch, Bell, van Bemmel, Bentum,
  Best, Bonafede, Breitling, Broderick, Brouw, Brüggen, Ciardi, Corstanje,
  de~Gasperin, de~Geus, Eislöffel, Engels, Falcke, Garrett, Grießmeier,
  Gunst, van Haarlem, Heald, Hoeft, Hörandel, Horneffer, Intema, Juette,
  Kuniyoshi, van Leeuwen, Loose, Maat, McFadden, McKay-Bukowski, McKean,
  Mulcahy, Munk, Pandey-Pommier, Polatidis, Reich, Röttgering, Rowlinson,
  Scaife, Schwarz, Steinmetz, Swinbank, Thoudam, Toribio, Vermeulen, Vocks, van
  Weeren, Wise, Yatawatta, \& Zarka}]{jackson_lbcs_2021}
Jackson, N., Tagore, A., Deller, A., {et~al.} 2021, Astronomy and Astrophysics,
  0, A86

\bibitem[{Kappes {et~al.}(2019)Kappes, Perucho, Kadler, Burd, Vega-García, \&
  Brüggen}]{kappes_lofar_2019}
Kappes, A., Perucho, M., Kadler, M., {et~al.} 2019, Astronomy and Astrophysics,
  631, A49

\bibitem[{McMullin {et~al.}(2007)McMullin, Waters, Schiebel, Young, \&
  Golap}]{mcmullin_casa_2007}
McMullin, J.~P., Waters, B., Schiebel, D., Young, W., \& Golap, K. 2007, in
  Astronomical {Data} {Analysis} {Software} and {Systems} {XVI}, ed. R.~A.
  Shaw, F.~Hill, \& D.~J. Bell, Vol. 376, 127

\bibitem[{{Mevius}(2018)}]{mevius_rmextract_2018}
{Mevius}, M. 2018, {RMextract: Ionospheric Faraday Rotation calculator}

\bibitem[{Mohan \& Rafferty(2015)}]{mohan_pybdsf_2015}
Mohan, N. \& Rafferty, D. 2015, Astrophysics Source Code Library, ascl:1502.007

\bibitem[{Moldón {et~al.}(2015)Moldón, Deller, Wucknitz, Jackson, Drabent,
  Carozzi, Conway, Kapińska, McKean, Morabito, Varenius, Zarka, Anderson,
  Asgekar, Avruch, Bell, Bentum, Bernardi, Best, Bîrzan, Bregman, Breitling,
  Broderick, Brüggen, Butcher, Carbone, Ciardi, de~Gasperin, de~Geus, Duscha,
  Eislöffel, Engels, Falcke, Fallows, Fender, Ferrari, Frieswijk, Garrett,
  Grießmeier, Gunst, Hamaker, Hassall, Heald, Hoeft, Juette, Karastergiou,
  Kondratiev, Kramer, Kuniyoshi, Kuper, Maat, Mann, Markoff, McFadden,
  McKay-Bukowski, Morganti, Munk, Norden, Offringa, Orru, Paas, Pandey-Pommier,
  Pizzo, Polatidis, Reich, Röttgering, Rowlinson, Scaife, Schwarz, Sluman,
  Smirnov, Stappers, Steinmetz, Tagger, Tang, Tasse, Thoudam, Toribio,
  Vermeulen, Vocks, van Weeren, White, Wise, Yatawatta, \&
  Zensus}]{moldon_lofar_2015}
Moldón, J., Deller, A.~T., Wucknitz, O., {et~al.} 2015, Astronomy and
  Astrophysics, 574, A73

\bibitem[{Morabito {et~al.}(2016)Morabito, Deller, Röttgering, Miley,
  Varenius, Shimwell, Moldón, Jackson, Morganti, van Weeren, \&
  Oonk}]{morabito_lofar_2016}
Morabito, L.~K., Deller, A.~T., Röttgering, H., {et~al.} 2016, Monthly Notices
  of the Royal Astronomical Society, 461, 2676

\bibitem[{Offringa(2010)}]{offringa_aoflagger_2010}
Offringa, A.~R. 2010, {AOFlagger}: {RFI} {Software}, published: Astrophysics
  Source Code Library

\bibitem[{{Offringa}(2016)}]{offringa_compression_2016}
{Offringa}, A.~R. 2016, \aap, 595, A99

\bibitem[{Offringa {et~al.}(2014)Offringa, McKinley, Hurley-Walker,
  {et~al.}}]{offringa_wsclean_2014}
Offringa, A.~R., McKinley, B., Hurley-Walker, {et~al.} 2014, MNRAS, 444, 606

\bibitem[{Offringa \& Smirnov(2017)}]{offringa_wsclean_2017}
Offringa, A.~R. \& Smirnov, O. 2017, MNRAS, 471, 301

\bibitem[{{Ram{\'\i}rez-Olivencia} {et~al.}(2018){Ram{\'\i}rez-Olivencia},
  {Varenius}, {P{\'e}rez-Torres}, {Alberdi}, {P{\'e}rez}, {Alonso-Herrero},
  {Deller}, {Herrero-Illana}, {Mold{\'o}n}, {Barcos-Mu{\~n}oz}, \&
  {Mart{\'\i}-Vidal}}]{ramirez_subarcsecond_2018}
{Ram{\'\i}rez-Olivencia}, N., {Varenius}, E., {P{\'e}rez-Torres}, M., {et~al.}
  2018, \aap, 610, L18

\bibitem[{{Roger} {et~al.}(1973){Roger}, {Costain}, \&
  {Bridle}}]{roger_low-frequency_1973}
{Roger}, R.~S., {Costain}, C.~H., \& {Bridle}, A.~H. 1973, \aj, 78, 1030

\bibitem[{{Schwab} \& {Cotton}(1983)}]{schwab_global_1983}
{Schwab}, F.~R. \& {Cotton}, W.~D. 1983, \aj, 88, 688

\bibitem[{Shepherd(1997)}]{shepherd_difmap_1997}
Shepherd, M.~C. 1997, 125, 77

\bibitem[{Shimwell {et~al.}(2017)Shimwell, Röttgering, Best, Williams,
  Dijkema, de~Gasperin, Hardcastle, Heald, Hoang, Horneffer, Intema, Mahony,
  Mandal, Mechev, Morabito, Oonk, Rafferty, Retana-Montenegro, Sabater, Tasse,
  van Weeren, Brüggen, Brunetti, Chyży, Conway, Haverkorn, Jackson, Jarvis,
  McKean, Miley, Morganti, White, Wise, van Bemmel, Beck, Brienza, Bonafede,
  Calistro~Rivera, Cassano, Clarke, Cseh, Deller, Drabent, van Driel, Engels,
  Falcke, Ferrari, Fröhlich, Garrett, Harwood, Heesen, Hoeft, Horellou,
  Israel, Kapińska, Kunert-Bajraszewska, McKay, Mohan, Orrú, Pizzo, Prandoni,
  Schwarz, Shulevski, Sipior, Smith, Sridhar, Steinmetz, Stroe, Varenius,
  van~der Werf, Zensus, \& Zwart}]{shimwell_lofar_2017}
Shimwell, T.~W., Röttgering, H. J.~A., Best, P.~N., {et~al.} 2017, Astronomy
  and Astrophysics, 598, A104

\bibitem[{Shimwell {et~al.}(2019)Shimwell, Tasse, Hardcastle, Mechev, Williams,
  Best, Röttgering, Callingham, Dijkema, de~Gasperin, Hoang, Hugo, Mirmont,
  Oonk, Prandoni, Rafferty, Sabater, Smirnov, van Weeren, White, Atemkeng,
  Bester, Bonnassieux, Brüggen, Brunetti, Chyży, Cochrane, Conway, Croston,
  Danezi, Duncan, Haverkorn, Heald, Iacobelli, Intema, Jackson, Jamrozy,
  Jarvis, Lakhoo, Mevius, Miley, Morabito, Morganti, Nisbet, Orrú, Perkins,
  Pizzo, Schrijvers, Smith, Vermeulen, Wise, Alegre, Bacon, van Bemmel,
  Beswick, Bonafede, Botteon, Bourke, Brienza, Calistro~Rivera, Cassano,
  Clarke, Conselice, Dettmar, Drabent, Dumba, Emig, Enßlin, Ferrari, Garrett,
  Génova-Santos, Goyal, Gürkan, Hale, Harwood, Heesen, Hoeft, Horellou,
  Jackson, Kokotanekov, Kondapally, Kunert-Bajraszewska, Mahatma, Mahony,
  Mandal, McKean, Merloni, Mingo, Miskolczi, Mooney, Nikiel-Wroczyński,
  O'Sullivan, Quinn, Reich, Roskowiński, Rowlinson, Savini, Saxena, Schwarz,
  Shulevski, Sridhar, Stacey, Urquhart, van~der Wiel, Varenius, Webster, \&
  Wilber}]{shimwell_lofar_2019}
Shimwell, T.~W., Tasse, C., Hardcastle, M.~J., {et~al.} 2019, Astronomy and
  Astrophysics, 622, A1

\bibitem[{Smirnov(2011)}]{smirnov_revisiting_2011}
Smirnov, O.~M. 2011, Astronomy and Astrophysics, 527, A106

\bibitem[{Smirnov \& Tasse(2015)}]{smirnov_radio_2015}
Smirnov, O.~M. \& Tasse, C. 2015, {\textbackslash}mnras, 449, 2668

\bibitem[{Tasse(2014)}]{Tasse_applying_2014}
Tasse, C. 2014, arXiv e-prints, 1410, arXiv:1410.8706

\bibitem[{Tasse {et~al.}(2017)Tasse, Hugo, Mirmont, Smirnov, Atemkeng, Bester,
  Hardcastle, Lakhoo, Perkins, \& Shimwell}]{tasse_faceting_2017}
Tasse, C., Hugo, B., Mirmont, M., {et~al.} 2017, ArXiv e-prints, 1712,
  arXiv:1712.02078

\bibitem[{Tasse {et~al.}(2018)Tasse, Hugo, Mirmont, Smirnov, Atemkeng, Bester,
  Hardcastle, Lakhoo, Perkins, \& Shimwell}]{tasse_faceting_2018}
Tasse, C., Hugo, B., Mirmont, M., {et~al.} 2018, Astronomy and Astrophysics,
  611, A87

\bibitem[{{Tasse} {et~al.}(2020){Tasse}, {Shimwell}, {Hardcastle},
  {O'Sullivan}, {van Weeren}, {Best}, {Bester}, {Hugo}, {Smirnov}, {Sabater},
  {Calistro-Rivera}, {de Gasperin}, {Morabito}, {R{\"o}ttgering}, {Williams},
  {Bonato}, {Bondi}, {Botteon}, {Br{\"u}ggen}, {Brunetti}, {Chy{\.z}y},
  {Garrett}, {G{\"u}rkan}, {Jarvis}, {Kondapally}, {Mandal}, {Prandoni},
  {Repetti}, {Retana-Montenegro}, {Schwarz}, {Shulevski}, \&
  {Wiaux}}]{tasse_lofar_2020}
{Tasse}, C., {Shimwell}, T., {Hardcastle}, M.~J., {et~al.} 2020, arXiv
  e-prints, arXiv:2011.08328

\bibitem[{{The HDF Group}(2000-2010)}]{hdf5}
{The HDF Group}. 2000-2010, {Hierarchical data format version 5}

\bibitem[{{van Bemmel} {et~al.}(2018){van Bemmel}, {Small}, {Kettenis},
  {Szomoru}, {Moellenbrock}, \& {Janssen}}]{van_bemmel_casa_2018}
{van Bemmel}, I., {Small}, D., {Kettenis}, M., {et~al.} 2018, in 14th European
  VLBI Network Symposium \& Users Meeting (EVN 2018), 79

\bibitem[{{van der Tol} {et~al.}(2007){van der Tol}, {Jeffs}, \& {van der
  Veen}}]{van_der_tol_self-calibration_2007}
{van der Tol}, S., {Jeffs}, B.~D., \& {van der Veen}, A.~J. 2007, IEEE
  Transactions on Signal Processing, 55, 4497

\bibitem[{{van Diepen} {et~al.}(2018){van Diepen}, {Dijkema}, \&
  {Offringa}}]{van_diepen_dppp_2018}
{van Diepen}, G., {Dijkema}, T.~J., \& {Offringa}, A. 2018, {DPPP: Default
  Pre-Processing Pipeline}

\bibitem[{van Haarlem {et~al.}(2013)van Haarlem, Wise, Gunst, Heald, McKean,
  Hessels, de~Bruyn, Nijboer, Swinbank, Fallows, Brentjens, Nelles, Beck,
  Falcke, Fender, Hörandel, Koopmans, Mann, Miley, Röttgering, Stappers,
  Wijers, Zaroubi, van~den Akker, Alexov, Anderson, Anderson, van Ardenne,
  Arts, Asgekar, Avruch, Batejat, Bähren, Bell, Bell, van Bemmel, Bennema,
  Bentum, Bernardi, Best, Bîrzan, Bonafede, Boonstra, Braun, Bregman,
  Breitling, van~de Brink, Broderick, Broekema, Brouw, Brüggen, Butcher, van
  Cappellen, Ciardi, Coenen, Conway, Coolen, Corstanje, Damstra, Davies,
  Deller, Dettmar, van Diepen, Dijkstra, Donker, Doorduin, Dromer, Drost, van
  Duin, Eislöffel, van Enst, Ferrari, Frieswijk, Gankema, Garrett,
  de~Gasperin, Gerbers, de~Geus, Griessmeier, Grit, Gruppen, Hamaker, Hassall,
  Hoeft, Holties, Horneffer, van~der Horst, van Houwelingen, Huijgen,
  Iacobelli, Intema, Jackson, Jelic, de~Jong, Juette, Kant, Karastergiou,
  Koers, Kollen, Kondratiev, Kooistra, Koopman, Koster, Kuniyoshi, Kramer,
  Kuper, Lambropoulos, Law, van Leeuwen, Lemaitre, Loose, Maat, Macario,
  Markoff, Masters, McFadden, McKay-Bukowski, Meijering, Meulman, Mevius,
  Middelberg, Millenaar, Miller-Jones, Mohan, Mol, Morawietz, Morganti,
  Mulcahy, Mulder, Munk, Nieuwenhuis, van Nieuwpoort, Noordam, Norden, Noutsos,
  Offringa, Olofsson, Omar, Orrú, Overeem, Paas, Pandey-Pommier, Pandey,
  Pizzo, Polatidis, Rafferty, Rawlings, Reich, de~Reijer, Reitsma, Renting,
  Riemers, Rol, Romein, Roosjen, Ruiter, Scaife, van~der Schaaf, Scheers,
  Schellart, Schoenmakers, Schoonderbeek, Serylak, Shulevski, Sluman, Smirnov,
  Sobey, Spreeuw, Steinmetz, Sterks, Stiepel, Stuurwold, Tagger, Tang, Tasse,
  Thomas, Thoudam, Toribio, van~der Tol, Usov, van Veelen, van~der Veen, ter
  Veen, Verbiest, Vermeulen, Vermaas, Vocks, Vogt, de~Vos, van~der Wal, van
  Weeren, Weggemans, Weltevrede, White, Wijnholds, Wilhelmsson, Wucknitz,
  Yatawatta, Zarka, Zensus, \& van Zwieten}]{van_haarlem_lofar_2013}
van Haarlem, M.~P., Wise, M.~W., Gunst, A.~W., {et~al.} 2013, Astronomy and
  Astrophysics, 556, A2

\bibitem[{van Weeren {et~al.}(2016)van Weeren, Williams, Hardcastle, Shimwell,
  Rafferty, Sabater, Heald, Sridhar, Dijkema, Brunetti, Brüggen,
  Andrade-Santos, Ogrean, Röttgering, Dawson, Forman, de~Gasperin, Jones,
  Miley, Rudnick, Sarazin, Bonafede, Best, Bîrzan, Cassano, Chyży, Croston,
  Ensslin, Ferrari, Hoeft, Horellou, Jarvis, Kraft, Mevius, Intema, Murray,
  Orrú, Pizzo, Simionescu, Stroe, van~der Tol, \&
  White}]{van_weeren_lofar_2016}
van Weeren, R.~J., Williams, W.~L., Hardcastle, M.~J., {et~al.} 2016, The
  Astrophysical Journal Supplement Series, 223, 2

\bibitem[{Varenius {et~al.}(2016)Varenius, Conway, Martí-Vidal, Aalto,
  Barcos-Muñoz, König, Pérez-Torres, Deller, Moldón, Gallagher, Yoast-Hull,
  Horellou, Morabito, Alberdi, Jackson, Beswick, Carozzi, Wucknitz, \&
  Ramírez-Olivencia}]{varenius_subarcsecond_2016}
Varenius, E., Conway, J.~E., Martí-Vidal, I., {et~al.} 2016, Astronomy and
  Astrophysics, 593, A86

\bibitem[{Varenius {et~al.}(2015)Varenius, Conway, Martí-Vidal, Beswick,
  Deller, Wucknitz, Jackson, Adebahr, Pérez-Torres, Chyży, Carozzi, Moldón,
  Aalto, Beck, Best, Dettmar, van Driel, Brunetti, Brüggen, Haverkorn, Heald,
  Horellou, Jarvis, Morabito, Miley, Röttgering, Toribio, \&
  White}]{varenius_subarcsecond_2015}
Varenius, E., Conway, J.~E., Martí-Vidal, I., {et~al.} 2015, Astronomy and
  Astrophysics, 574, A114

\bibitem[{{Williams} {et~al.}(2019){Williams}, {Hardcastle}, {Best}, {Sabater},
  {Croston}, {Duncan}, {Shimwell}, {R{\"o}ttgering}, {Nisbet}, {G{\"u}rkan},
  {Alegre}, {Cochrane}, {Goyal}, {Hale}, {Jackson}, {Jamrozy}, {Kondapally},
  {Kunert-Bajraszewska}, {Mahatma}, {Mingo}, {Morabito}, {Prandoni},
  {Roskowinski}, {Shulevski}, {Smith}, {Tasse}, {Urquhart}, {Webster}, {White},
  {Beswick}, {Callingham}, {Chy{\.z}y}, {de Gasperin}, {Harwood}, {Hoeft},
  {Iacobelli}, {McKean}, {Mechev}, {Miley}, {Schwarz}, \& {van
  Weeren}}]{williams_lofar_2019}
{Williams}, W.~L., {Hardcastle}, M.~J., {Best}, P.~N., {et~al.} 2019, \aap,
  622, A2

\end{thebibliography}
 
\begin{appendix}
\section{Pipeline profile}
\label{app:profile}

In this appendix we provide a basic pipeline profile that describes the number of core hours necessary to run the pipeline. The number of core hours is used as every pipeline is parallelised to make full use of the available computing resources. Although \prefac\ and \ddfp\ are described in detail elsewhere, we also provide a brief summary for the reader. The \prefac\ calibrator processing required 245 core hours. In practice, this was $\sim$3 hours when executed on a single compute node with 192 GB RAM and two Intel Xeon Gold 6130 processors, which have 16 cores each and run at 2.1 GHz. The \prefac\ target processing took 1,444 core hours. In practice, this was just under 30 hours to process the full bandwidth when executed on a single compute node with 192 GB RAM and two Intel Xeon Gold 6130 processors, which have 16 cores each and run at 2.1 GHz. The \ddfp\ target processing took 5,389 core hours. In practice, this was $\sim$7 days for the full bandwidth of an eight hour pointing when executed on a single compute node with 192 GB RAM and two Intel Xeon Gold 6130 processors, which have 16 cores each and run at 2.1 GHz. 
The total pre-procesing time necessary is therefore 7,078 core hours. 

\begin{figure}
\includegraphics[width=0.5\textwidth,clip,trim=0.6cm 0cm 0cm 0cm]{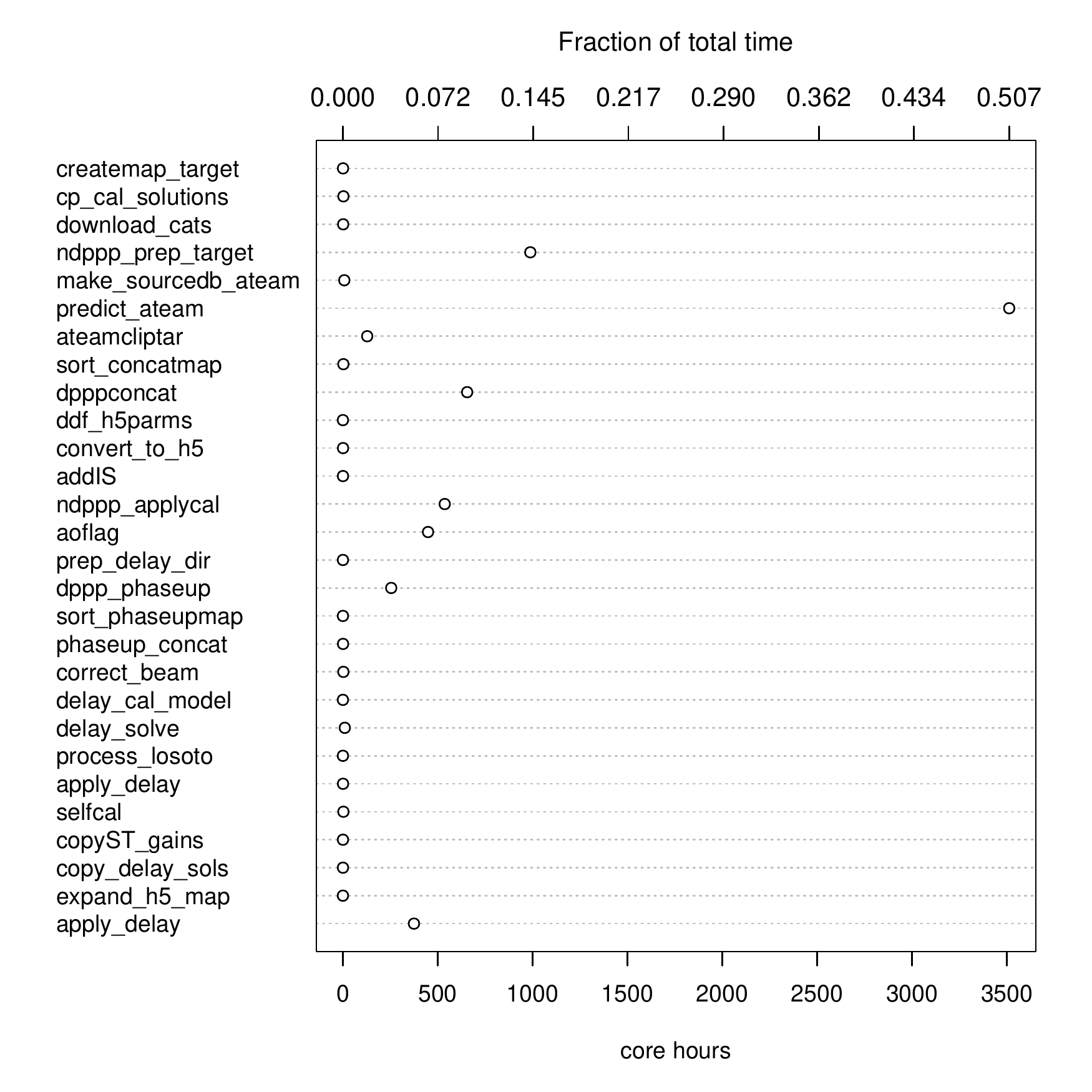}
\caption{\label{fig:steps} Dot chart showing the number of core hours (bottom axis) and fraction of total time (top axis) for each step.}
\end{figure}

To profile the \lofvlbi\ pipeline, we make use of the {\tt genericpipeline} logs output that are tagged with the system time. For the {\tt Delay-Calibration} step, we collected the start and end times for each individual step, found the average time per subband or band (depending on the step), multiplied the average time by the number of times the step was run (e.g. average time per subband $\times$ number of subbands) and finally factored in the number of cores used to calculate the core hours. These are listed in Table \ref{tab:steps} and shown visually in Figure~\ref{fig:steps}. Only steps that take longer than 1 second are recorded. Many steps are less than 1 core hour, with only a few steps dominating a large fraction of the total time. In particular, over half of the time is spent in the {\tt predict\_ateam} step. 

\begin{table}
\caption{\label{tab:steps}Core hours for each step in the pipeline. Only steps with more than 1 second of total time are listed.}
\begin{tabular}{lr}
Step & CPU hours \\[2pt] \hline \\[-6pt] 
createmap\_target & 0.60 \\ 
cp\_cal\_solutions & 2.65 \\ 
download\_cats & 1.33 \\ 
ndppp\_prep\_target & 987.77 \\ 
make\_sourcedb\_ateam & 7.29 \\ 
predict\_ateam & 3511.72 \\ 
ateamcliptar & 128.03 \\ 
sort\_concatmap & 2.43 \\ 
dpppconcat & 654.83 \\ 
ddf\_h5parms & 0.0067 \\ 
convert\_to\_h5 & 0.73 \\ 
addIS & 0.31 \\ 
ndppp\_applycal & 536.60 \\ 
aoflag & 448.68 \\ 
prep\_delay\_dir & 0.18 \\ 
dppp\_phaseup & 254.61 \\ 
sort\_phaseupmap & 0.013 \\ 
phaseup\_concat & 0.65 \\ 
correct\_beam & 2.47 \\ 
delay\_cal\_model & 0.01 \\ 
delay\_solve & 9.87 \\ 
process\_losoto & 0.03 \\ 
apply\_delay & 0.60 \\ 
selfcal & 2.52 \\ 
copyST\_gains & 0.28 \\ 
copy\_delay\_sols & 0.33 \\ 
expand\_h5\_map & 0.0033 \\ 
apply\_delay & 374.47 \\ 
\hline \\[-6pt] 
 \textbf{Total} & \textbf{ 6929.03 } \\ 
\end{tabular}
 \end{table}

This covers the total number of core hours for all of the common steps for individual directions. From here, sources are split out and processed independently. This can either be done with {\tt Split-Directions} or manually for further parallelisation. Here we report the total time per source, which is similar for both cases. Each source requires splitting out the direction twice: once with combining the core stations into ST001 before filtering them out, and once keeping the core stations (for the final imaging). Creating a single direction with ST001 and correcting the beam takes $\sim$51 core hours, while the same but keeping the core stations takes $\sim$142 core hours. Self-calibration of a source takes between $\sim$30 to $\sim$60 core hours, depending on how many cycles one uses. Final imaging takes around 6 core hours. Overall the time to process one direction completely is $\sim$230 to 260 core hours. 

The total estimated time to run the pipeline is therefore $\sim$7,000 core hours + (250 core hours)$\times$(number of sources). 

\end{appendix}

\end{document}